\documentclass[final,pdftex,11pt]{ectaart}

\RequirePackage[OT1]{fontenc} \RequirePackage{amsthm}
\RequirePackage[cmex10]{amsmath}\RequirePackage{natbib}
\RequirePackage[colorlinks,citecolor=blue,urlcolor=blue]{hyperref}
\RequirePackage{hypernat}

\usepackage{amsmath}
\usepackage{amsfonts}
\usepackage{amssymb}
\usepackage{graphicx}

\startlocaldefs \numberwithin{equation}{section}
\theoremstyle{plain}

\newtheorem{theorem}{Theorem}[section]
\newtheorem{corollary}{Corollary}[section]
\newtheorem{proposition}{Proposition}[section]
\newtheorem{definition}{Definition}[section]

\newtheorem{remark}{Remark}[section]
\endlocaldefs

\begin{document}

\begin{frontmatter}
\title{Robust valuation and risk measurement under model uncertainty \protect}
\runtitle{mean-volatility uncertainty}
\begin{aug}
\author{\fnms{Yuhong} \snm{Xu}\ead[label=e2]{yuhong.xu@hotmail.com}}
\runauthor{Y. Xu}
\address{The author would like to thank the supports of Shanghai Key Laboratory of Financial Information
Technology (SUFE) and the collaborative innovation center for quantitative
calculation and control of financial risk and the Project 111 (No. B12023). The author would also like to show many thanks to Prof. S. Peng and Dr. Xinpeng Li and other seminar participants of nonlinear expectation at Shandong university for helpful suggestions and particular thanks to Dr. Antoine Jacquier in Imperial College for discussions on the notion of P\&L.}
\address{ Address correspondence to Yuhong Xu, Center for Financial engineering, Soochow University,
Suzhou 215006, P.R.China.
\textbf{E-mail}: 
\printead{e2}}
\end{aug}
Soochow University \vspace{4mm}
\begin{abstract}
Model uncertainty is a type of inevitable financial risk. Mistakes on the
choice of pricing model may cause great financial losses. In this
paper we investigate financial markets with mean-volatility
uncertainty. Models for stock markets and option markets with
uncertain prior distribution are established by Peng's G-stochastic
calculus. The process of stock price is described by generalized
geometric G-Brownian motion in which the mean uncertainty may move together with or regardless of
the volatility uncertainty. On the hedging market, the upper price of an (exotic) option is derived following the Black-Scholes-Barenblatt equation. It is interesting that the corresponding Barenblatt equation does not depend on the risk preference
of investors and the mean-uncertainty of underlying stocks. Hence under
some appropriate sublinear expectation, neither the risk preference of
investors nor the mean-uncertainty of underlying stocks pose effects on
our super and subhedging strategies. Appropriate definitions of arbitrage for super and sub-hedging strategies
are presented such that  the super and sub-hedging prices are reasonable. Especially the condition of arbitrage for
sub-hedging strategy fills the gap of the theory of arbitrage under model uncertainty. Finally we show that the term $K$ of finite-variance arising in the super-hedging strategy is interpreted as the max Profit\&Loss of being short a delta-hedged option. The ask-bid spread is in fact the
accumulation of summation of the superhedging $P\&L$ and the subhedging $P\&L $.
\end{abstract}
\end{frontmatter}

\section{Introduction}

\label{v.sect1}

Mathematical models have come to play an important role in pricing and
hedging derivative instruments since Black and Scholes' seminal work (%
\textrm{\cite{bs}}). The Black-Scholes option pricing formula has been used
extensively, even to evaluate options whose underlying asset (e.g. the
stock) is known to not satisfy the Black-Scholes hypothesis of a constant
volatility. We go about our work as if we are correct, we often treat
parameters as if we think they are. And yet in the strict sense of word we
do not know how much we do not know. Unknown parameters, typically, mean and
volatility uncertainty lead to model risk\footnote{%
Model risk is the risk of error in our estimated risk measure due to
inadequacies in our risk models (\textrm{\cite{dowd}}). Model uncertainty
leads to a kind of model risk. Ambiguity on volatility is a typical case of
model uncertainty.} or model uncertainty. Model risk is an inescapable
consequence of model use. It is often hidden or glossed over and is often
overlooked. A failure to consider model risk can lead a firm to disaster,
and sometimes has, as pointed in \textrm{\cite{cont}}.

A typical case of model risk is the choice of probabilistic models. Often a
decision maker or a risk manager is not able to attribute a precise
probability to future outcomes. This situation has been called uncertainty
by \textrm{\cite{knight}}. Knight uncertainty sometimes is used to designate
the situation where probabilities are unknown. Alternatively, we speak of
\textit{ambiguity} when we are facing several possible specifications $%
P_{1},P_{2},\ldots $ for probabilities on future outcomes (\textrm{\cite%
{epstein}}). Ambiguity aversion has shown to have important consequences in
macroeconomics (\textrm{\cite{hst}}) and for price behavior in capital
markets (\textrm{\cite{ce, ew,rz}}). In this circumstance, fair option
values and perfectly replicating hedges cannot be determined with certainty.
The existence of volatility risk in derivative trading is a concrete
manifestation of market incompleteness.

The problem of model uncertainty has long been recognized in economics and
finance. \textrm{\cite{dw}} studied a single period portfolio choice problem
employing the uncertainty averse preference model developed by \textrm{\cite%
{schmdl}}. \textrm{\cite{ew94}} and\textrm{\ \cite{ce}} studied the
implications for equilibrium asset prices in the representative agent
economics. Cash-subadditive risk measures with interest ambiguity was
studied in \textrm{\cite{er}}. \textrm{\cite{xu10} }investigated
multidimensional risk measures under multiple priors. See also \textrm{\cite%
{ew}, \cite{gd},} \textrm{\cite{rd09} }and references therein for more
papers on model uncertainty and multi-prior model. We do not list them all
here. Note that in existing works on model uncertainty (\textrm{\cite{ce,
ew,gd,er,xu10}}), all probability measures\textrm{\ }$P\in \mathcal{P}$ are
assumed to be equivalent to a reference probability $P_{0}$. This technical
requirement is actually quite restrictive: it means all model agree on a
universe scenario and only differ on their probabilities. An example of
diffusion model with uncertain volatility (\textrm{\cite{alp,lyons,cont}})
does not verify this hypothesis. Recent explorations include \textrm{\cite%
{vorbrink}}, \textrm{\cite{ns},} \textrm{\cite{ej,ej13} }and \textrm{\cite%
{md12,ms12,empy14,empsy14}}.

Volatility of a financial market is difficult to predict. Although we have
lots of historical data within hand, the volatility might move as large as
she wants and seems to be quite sensitive to new information. One could
approximate short-period volatility but never the long-term one. There are
too many factors determining volatility. Sometimes we assume that the
volatility is driven by stochastic elements, e.g. itself is a diffusion
process. Such a model is called stochastic volatility model(\textrm{\cite%
{heston}}). It often has several parameters which can be chosen either to
fit historical data or calibrate to the market.

A robust choice to the problem of modeling the unknown volatility is to
treat it as uncertain as it actually is. We just stand on two bounds $%
\underline{\sigma }$ and $\overline{\sigma }$ to deduce prices representing
worst-case scenario and best-case scenario respectively. The interval [$%
\underline{\sigma },\overline{\sigma }$] characterizes the uncertain level
of volatility. Larger interval, larger fluctuation of volatility. Also this
interval depends on investor's preference or aversion of risk. A
conservative investor may establish a large interval and choose the minimal
superstrategy. However a too large interval yields such a high superstrategy
that it is meaningless.

We now recall the uncertain volatility model introduced in \textrm{\cite{alp}%
}. For simplicity, we only consider derivative securities based on a single
liquidly traded stock which pays no dividends over the contract's lifetime.
The paths followed by future stock prices are assumed to be It\^{o} process,
\begin{equation}
dS_{t}=S_{t}\left( u_{t}dt+\sigma _{t}dW_{t}\right) .
\end{equation}%
where $\left( u_{t}\right) $ and $\left( \sigma _{t}\right) $ are adapted
processes such that%
\begin{equation*}
\underline{\sigma }\leq \sigma _{t}\leq \overline{\sigma },
\end{equation*}%
where $\left( W_{t}\right) $ is the standard Brownian motion under a given
probability space $(\Omega ,\mathcal{F},P)$. The constants $\overline{\sigma
}$ and $\underline{\sigma }\ $represent upper and lower bounds of the
volatility that should be input to the model according to the investor's
expectation and uncertainty about future price fluctuations. These two
bounds could be statistically obtained from peaks of volatility in
historical stock or option-implied volatilities. They can be viewed as
determining a confidence interval for future volatility values, as pointed
in \textrm{\cite{alp}.}

Note that two different volatility processes will typically yield mutually
singular probability measures on the set of possible paths. So volatility
ambiguity leads to model uncertainty with a set of risk-neutral
probabilities $\mathcal{P}$, each of them corresponding to a volatility
process with value at each time in $\left[ \underline{\sigma },\overline{%
\sigma }\right] $. Naturally we look for the cheapest superhedging price at
which we can sell and manage an option in such environment. A convenient
framework is the stochastic control framework, in which the managing
volatility is interpreted as a control variable. It turns out that the value
function in such an optimal control will yield the cheapest superstrategy
price. Nevertheless, the connection between superstrategy problem and
stochastic control is not that obvious. Recall that a stochastic control
problem is to maximize an expectation over a set of processes, whereas a
superstrategy is over a set of probabilities, i.e., $\underset{P\in \mathcal{%
P}}{\sup }{\normalsize E}_{P}$. This issue is avoided in \textrm{\cite{alp},}
handled partially in \textrm{\cite{lyons}}, and more formally in \textrm{%
\cite{martini}} and \textrm{\cite{frey}}. A significant progress toward a
general framework is available in \textrm{\cite{dm}}, which can be viewed as
a quasi-sure stochastic analysis. See also \textrm{\cite{stza}} addressing
conditioning or updating which is a crucial ingredient in modeling dynamic
pricing.

\textrm{\cite{p06,p07b,p08}} established a path analysis, called
G-stochastic analysis, which extends the classical Wiener analysis to a
framework of sublinear expectation on events field $\Omega =C_{0}([0,+\infty
),\mathbf{R}^{d})$, the space of all $\mathbf{R}^{d}$-valued continuous
paths $(\omega _{t})_{t\in R^{+}}$ with $\omega _{0}=0$, equipped with a
uniform norm on compact subspaces. Notions such as G-normal distribution,
G-Brownian motion, G-expectation were introduced (see Appendix \ref{v.appA}
or Peng's review paper, \textrm{\cite{p09}}, and summative book, \textrm{%
\cite{p10a}}).

The representation for G-expectation (\textrm{\cite{hp}}),%
\begin{equation*}
\mathbb{E}\left[ \cdot \right] =\underset{P\in \mathcal{P}}{\sup }%
{\normalsize E}_{P}\left[ \cdot \right]
\end{equation*}%
tells us that G-expectation induces a set of probabilities $\mathcal{P}$
naturally.\footnote{$\mathcal{P}$ is a weekly compact set. Recently \textrm{%
\cite{bk}} showed that there is a numerable weakly relatively compact set $%
\left\{ P_{n},n\in N\right\} \subset \mathcal{P}$ such that the above
representation still holds.} It is shown that $\left( B_{t}\right) $ is a
martingale under every $P\in \mathcal{P}$ (\textrm{\cite{stzc,ns}}) and
there exists a unique adapted process $(\sigma _{t}^{p})$ such that $%
\underline{\sigma }^{2}\leq \left\vert \sigma _{t}^{p}\right\vert ^{2}\leq
\overline{\sigma }^{2}$, $a.a.$\footnote{%
a.a.: almost all; a.s.: almost surely; a.e.: almost everywhere.} $t$, $P$-$%
a.s.$ and
\begin{equation*}
B_{t}=\int_{0}^{t}\sigma _{s}^{p}dW_{s}^{P},\forall t\geq 0,P\text{-}a.s.
\end{equation*}%
where $(W_{t}^{P})$ is a standard $E_{P}-$Brownian motion. Therefore an
interesting phenomenon comes up: the quadratic variance of $\left(
B_{t}\right) $ under any $P\in \mathcal{P}$,
\begin{equation*}
\left\langle B\right\rangle _{t}^{P}=\int_{0}^{t}\left\vert \sigma
_{s}^{p}\right\vert ^{2}ds,\forall t\geq 0,P-a.s.
\end{equation*}%
is no longer a deterministic function of time $t$.

All results in G-stochastic analysis work in a model-free way: They hold
under all probabilities $P\in \mathcal{P}$ or quasi-surely(\textit{q.s.}),
i.e. a property holds outside a polar set $A$ with $P(A)=0$ for all $P\in
\mathcal{P}$. As pointed in Peng's ICM\footnote{%
International Congress of Mathematicians} lecture (\textrm{\cite{p10b}}),
G-expectation may appear as a natural candidate to measure volatility risk.
In this direction, initial work has been done by \textrm{\cite{vorbrink} i}n
which the main focus is on the no-arbitrage argument. Based on the work of
\textrm{\cite{kk}}, Vorbrink adapted the notion of absence of arbitrage from
the market with constraints on portfolio choice to the framework of
uncertain volatility model. The risk premium of portfolio is not considered
when modeling the wealth process. Thus technical difficulties to change the
subjective risk preference to a risk-neutral world are avoided. More
recently, recursive utility is studied by \textrm{\cite{ej,ej13}}
accommodating mean and volatility ambiguity. They applied the model to a
representative agent endowment economy to study equilibrium asset returns in
both Arrow-Debreu style and sequential Radner-style economies. \textrm{\cite%
{md12}} presents an equilibrium model for two-price economics in which the
market clearing condition is defined. See also \textrm{\cite{ms12, empy14,
empsy14}} for related results using the theory of G-expectation.

The present paper considers mean-volatility uncertainty simultaneously. As
pointed later in the next section, mean-uncertainty occurs often with
volatility uncertainty. The stock price is modeled as a generalized
geometric G-Brownian motion in which mean-uncertainty may move without
regarding to the volatility-uncertainty. Section \ref{v.sect3} derives the
superhedging PDEs for both state-dependent and discrete-path-dependent
options. What is interesting is that, neither the preference of investors
nor the mean-uncertainty appear in the superhedging PDEs, which demonstrates
that risk-neutral measures exist indeed in such ambiguous environment.
Section \ref{v.sect4} extends the classical Black-Scholes-Merton model to
the uncertain volatility case. A superhedging strategy is just a solution of
a backward stochastic differential equation driven by G-Brownian motion
(G-BSDE for short). It is shown that the solution is the minimal
superstrategy with no-arbitrage. In particular, conditions of arbitrage for
substrategy are given which are essentially different from Vorbink's work.
The finite-variance term $K$ is interpreted as Profit\&Loss (P\&L\footnote{%
See \textrm{\cite{mj}}, \textrm{\cite{js}} for the notion of P\&L.} for
short) of an investor. Recall of Peng's G-stochastic analysis and some
technical results are arranged in the appendix.

\section{Mean-volatility uncertainty of stocks}

\label{v.sect2}

\subsection{Volatility-uncertainty brings mean-uncertainty}

\label{v.sect2.3}We assume the price process of a stock satisfies the
following linear stochastic differential equation (SDE for short):%
\begin{equation*}
dS_{t}=S_{t}\left( \mu dt+dB_{t}\right) ,
\end{equation*}%
where $\left( B_{t}\right) $ is a G-Brownian motion\footnote{%
For convenience of writing, in the following of the paper, we will always
consider models driven by one dimensional $G$-Brownian motion.}. Define the
continuously compounded rate of return per annum realized between 0 and T as
$\zeta $. It follows that
\begin{equation*}
S_{T}=S_{0}e^{\zeta T}
\end{equation*}%
and
\begin{equation*}
\zeta =\frac{1}{T}\ln \frac{S_{T}}{S_{0}}=\frac{1}{T}\left\{ B_{T}+\mu T-%
\frac{1}{2}\left\langle B\right\rangle _{T}\right\} .
\end{equation*}%
So the mean of expected continuously compounded rate of return will
fluctuate within $[\mu -\frac{1}{2}\overline{\sigma }^{2},\mu -\frac{1}{2}%
\underline{\sigma }^{2}]$. We do not consider any ambiguity of stock
appreciation. However the mean or the expected rate of return is uncertain. $%
S_{t}$ is not a symmetric random variable at each time t, since we do not
necessarily have $\mathbb{E}[S_{t}]=-\mathbb{E}[-S_{t}]$.

Volatility ambiguity leads to model uncertainty, i.e., multi-prior model.
Naturally the expected value of stock price $S_{t}$ may be ambiguous under a
set of probabilistic models. This paper will take into account
mean-volatility uncertainty simultaneously by Peng's G-stochastic analysis.

\subsection{The process for stock prices}

\label{v.sect2.4}In the classical Black-Scholes-Merton option-pricing model,
the price process of a stock is assumed to be It\^{o} process
\begin{equation}
dS_{t}=S_{t}\left( \mu _{t}dt+\sigma _{t}dW_{t}\right) ,  \label{v.2.1}
\end{equation}%
where $W$ is the standard Brownian motion under a given linear probability
space $(\Omega ,\mathcal{F},P)$; $\sigma _{t}$ is the volatility of the
stock price; $\mu _{t}$ is the expected rate of return.

An application of It\^{o} formula yields%
\begin{equation*}
S_{t}=S_{0}\exp\left\{ \int_{0}^{t}\sigma_{s}dW_{s}+\int_{0}^{t}(\mu _{s}-%
\frac{1}{2}\sigma_{s}^{2})ds\right\} .
\end{equation*}
which is called geometric Brownian motion.

We now consider a stock market with mean-uncertainty and
volatility-uncertainty together. We do not have confidence in which
direction the expected rate $\mu $ of return and the volatility $\sigma $
will move or even their distribution in future but they are sure to change
within $[\underline{\mu },\overline{\mu }]$ and $[\underline{\sigma },%
\overline{\sigma }]$. This uncertain model could be described by
finite-variance $G_{[\underline{\mu },\overline{\mu }]}$-Brownian motion and
zero-mean $G_{[\underline{\sigma }^{2},\overline{\sigma }^{2}]}$-Brownian
motion together. Let $\left( \beta _{t}\right) $ be a finite-variance $G_{[%
\underline{\mu },\overline{\mu }]}$-Brownian motion and $\left( B_{t}\right)
$ a zero-mean $G_{[\underline{\sigma }^{2},\overline{\sigma }^{2}]}$%
-Brownian motion under a given sublinear expectation $\mathbb{E}$. Then
model \eqref{v.2.1} could be rewritten as%
\begin{equation}
dS_{t}=S_{t}\left( d\beta _{t}+dB_{t}\right) .  \label{v.2.2}
\end{equation}

However, we prefer the following modification about the expected rate of
returns: let $r$ be the riskless interest rate. If $\mu $ varies in $[%
\underline{\mu },\overline{\mu }]$, then $\mu -r$ varies in $[\underline{\mu
}-r,\overline{\mu }-r]$. Let $\left( \beta _{t}\right) $ be a
finite-variance $G_{[\underline{\mu }-r,\overline{\mu }-r]}$-Brownian
motion, then \eqref{v.2.2} is in form of
\begin{equation}
dS_{t}=S_{t}\left( rdt+d\beta _{t}+dB_{t}\right) .  \label{v.2.3}
\end{equation}

It is important to keep in mind that we do not assume a risk-neutral world
in advance in model \eqref{v.2.3}. Of course we will see later that there
does exist a risk-neutral world in which even the uncertainty of expected
returns does not influence our super and sub-hedging strategies.

Particularly taking $\beta _{t}=\left\langle B\right\rangle _{t}$ means that
the expected returns and volatility move together. See \textrm{\cite{xsz}},
\textrm{\cite{osuka}} and \textrm{\cite{bei}}, they consider
mean-uncertainty of this type. An example is referred in \textrm{\cite{ej}}
Example 2.4 and\textrm{\ \cite{es}} Section 3.1.2\textrm{\ }by specifying

\begin{equation*}
\mu=\underline{\mu}+z,\sigma^{2}=\underline{\sigma}^{2}+\frac{2z}{\gamma},
\end{equation*}
where $0\leq z\leq\overline{z}$ and $\underline{\mu},\underline{\sigma },%
\overline{z}$ and $\gamma$ are fixed and known parameters, which means that $%
\mu=\underline{\mu}+\frac{\gamma}{2}\left( \sigma^{2}-\underline{\sigma}%
^{2}\right) $ and yields%
\begin{equation}
dS_{t}=S_{t}\left[ \left( \underline{\mu}-\frac{\gamma}{2}\underline{\sigma }%
^{2}\right) dt+\frac{\gamma}{2}d\left\langle B\right\rangle _{t}+dB_{t}%
\right]
\end{equation}
or equivalently

\begin{equation*}
dS_{t}=S_{t}\left( rdt+d\beta _{t}+dB_{t}\right) ,where\ \beta _{t}=\left(
\underline{\mu }-r-\frac{\gamma }{2}\underline{\sigma }^{2}\right) t+\frac{%
\gamma }{2}\left\langle B\right\rangle _{t}.
\end{equation*}

\textrm{\cite{id}} showed that such models exist when agents receive bad
news of ambiguous precision since bad news lowers both the conditional mean
and the conditional variance of returns.

\subsection{Approximate evaluation for stocks}

\label{v.sect2.6}The classical price process of stock yields
\begin{equation*}
\ln S_{t}\sim \mathcal{N}\left( \ln S_{0}+(r-\frac{1}{2}\sigma ^{2})T,\
\sigma \sqrt{T}\right)
\end{equation*}%
(see \textrm{\cite{hull}}) where $\mathcal{N}$ is the distribution function
of normal distribution. There is a 95\% probability that a normally
distributed variable has a value with 1.96 standard deviation of its mean.
Hence, with 95\% confidence under a single $P$ we have
\begin{equation*}
\ln S_{0}+(r-\frac{1}{2}\sigma ^{2})T-1.96\sigma \sqrt{T}<\ln S_{T}<\ln
S_{0}+(r-\frac{1}{2}\sigma ^{2})T+1.96\sigma \sqrt{T}.
\end{equation*}

Typical values of the volatility of a stock are in the range of 20\% to 40\%
per anum and usually we take $T\leq 1$. If We define
\begin{equation*}
f_{1}(\sigma )=-\frac{1}{2}\sigma ^{2}T-1.96\sigma \sqrt{T},
\end{equation*}%
\begin{equation*}
f_{2}(\sigma )=-\frac{1}{2}\sigma ^{2}T+1.96\sigma \sqrt{T},
\end{equation*}%
it is easy to check that when $\sigma $ takes values in $\left[ \underline{%
\sigma },\overline{\sigma }\right] \subseteq \lbrack 0.2,0.4]$, the function
$f_{1}$ is decreasing and $f_{2}$ increasing. If we take the maximum of
volatility $\overline{\sigma }$, we have that

$\bullet $ For any $P\in \mathcal{P}$, with at least 95\% confidence we have
\begin{equation*}
\ln S_{0}+(r-\frac{1}{2}\overline{\sigma }^{2})T-1.96\overline{\sigma }\sqrt{%
T}<\ln S_{T}<\ln S_{0}+(r-\frac{1}{2}\overline{\sigma }^{2})T+1.96\overline{%
\sigma }\sqrt{T}.
\end{equation*}

\section{Risk-neutral \& mean-certain valuation}

\label{v.sect3}This section derives the superhedging PDEs for both
state-dependent and discrete-path-dependent options, which shows the
existence of a risk-neutral \& mean-certain world in which all investors are
hedging without the influence of risk preference and mean-uncertainty.

\subsection{State-dependent payoffs}

\label{v.sect3.1}The Black-Scholes equation is derived for state-dependent
European options. Now we derive the superhedging PDE within this framework
which is easy to understand and comparable with Black-Scholes-Merton's
model. We assume the price process of the stock satisfies the following SDE:%
\begin{equation}
dS_{t}=S_{t}\left( \mu _{t}dt+\sigma _{t}dW_{t}\right) ,  \label{v.3.1}
\end{equation}%
where $W$ is the standard Brownian motion under a given linear probability
space $(\Omega ,\mathcal{F},P)$; $\mu $ is the expected rate of return
varying in $[\underline{\mu },\overline{\mu }]$; $\sigma $ is the volatility
of the stock price varying in $\left[ \underline{\sigma },\overline{\sigma }%
\right] $. $\left( \mu _{t}\right) $, $\left( \sigma _{t}\right) $ and the
riskless interest rate $\left( r_{t}\right) $ are assumed to be
deterministic functions of $t$. Note that we do not assume any relation
between $\mu $ and $\sigma $. The mean uncertainty may move together with or
regardless of the volatility uncertainty, while in \textrm{\cite{xsz}},
\textrm{\cite{osuka}} and \textrm{\cite{bei}}, they in fact consider the
case $\mu _{t}=\sigma _{t}^{2}$. Note also that we do not assume in advance
a risk-neutral world which is different from \textrm{\cite{jxrl}} and
\textrm{\cite{meyer}.}

Let $V(t,S_{t})$ be the price of the option with payoff $\Phi (S_{T})$,
where $V$ and $\Phi $ are both deterministic function. Assume also that $%
V\in C^{1,2}\left( \left[ 0,T\right] \times \mathbf{R}\right) $.\footnote{$%
C^{j,k}\left( \left( 0,T\right) \times \mathbf{R}\right) $ denotes the set
of functions defined on $\left( 0,T\right) \times \mathbf{R}$ which are $j$
times differentiable in $t\in \left( 0,T\right) $ and $k$ times
differentiable in $x\in \mathbf{R}$ such that all these derivatives are
continuous.} By It\^{o}'s formula,

\begin{equation}
dV(t,S_{t})=\frac{\partial V}{\partial t}dt+S_{t}\frac{\partial V}{\partial S%
}\left( \mu _{t}dt+\sigma _{t}dW_{t}\right) +\frac{1}{2}\sigma _{t}^{%
{\normalsize 2}}S_{t}^{{\normalsize 2}}\frac{\partial ^{2}V}{\partial S^{2}}%
dt.  \label{v.3.2}
\end{equation}%
The discrete versions of equations \eqref{v.3.1} and \eqref{v.3.2} are
\begin{equation}
\Delta S_{t}=S_{t}\left( \mu _{t}\Delta t+\sigma _{t}\Delta W_{t}\right)
\label{v.3.3}
\end{equation}%
and%
\begin{equation}
\Delta V_{t}=\frac{\partial V}{\partial t}\Delta t+S_{t}\frac{\partial V}{%
\partial S}\left( \mu _{t}\Delta t+\sigma _{t}\Delta W_{t}\right) +\frac{1}{2%
}\sigma _{t}^{{\normalsize 2}}S_{t}^{{\normalsize 2}}\frac{\partial ^{2}V}{%
\partial S^{2}}\Delta t.  \label{v.3.4}
\end{equation}

For a delta-hedging portfolio $\Pi $, the holder of this portfolio is short
one derivative and long an amount $\frac{\partial V}{\partial S}$ of shares
of stocks and $\left( V-\frac{\partial V}{\partial S}S\right) $ cash left in
a bank account. Then the P\&L variance of the portfolio is
\begin{equation}
\Delta \Pi _{t}=\frac{\partial V}{\partial S}\Delta S_{t}-\Delta
V_{t}+\left( V_{t}-\frac{\partial V}{\partial S}S_{t}\right) r_{t}\Delta t.
\label{v.3.5}
\end{equation}%
The first part corresponds to the stock price movements, of which we hold $%
\frac{\partial V}{\partial S}$ units, the second one to the price variation
of the option, and the third part is the risk-free return of the amount of
cash to make the portfolio have zero value. Now, substituting equations %
\eqref{v.3.3} and \eqref{v.3.4} into \eqref{v.3.5} yields%
\begin{equation*}
\Delta \Pi _{t}=-\frac{\partial V}{\partial t}\Delta t-\frac{1}{2}\sigma
_{t}^{{\normalsize 2}}S_{t}^{{\normalsize 2}}\frac{\partial ^{2}V}{\partial
S^{2}}\Delta t+\left( V_{t}-\frac{\partial V}{\partial S}S_{t}\right)
r_{t}\Delta t.
\end{equation*}%
Observe that neither the random noise nor the stock appreciation arise in $%
\Delta \Pi $ explicitly. If the managing volatility\footnote{%
The managing volatility is the volatility at which the option is sold.} of
the option coincides with the realised volatility of stocks, of course, by
the principle of no-arbitrage, $\Delta \Pi _{t}=0$. However it is unclear
which is the realised volatility. The seller of the option wishes to find a
cheapest managing policy yielding a non-negative P\&L, at least no loss.
More precisely, we want to have \
\begin{equation*}
\underset{\underline{\sigma }\leq \sigma _{t}\leq \overline{\sigma }}{\inf }%
\Delta \Pi _{t}=0.
\end{equation*}%
Consequently, we deduce that%
\begin{eqnarray}
\frac{\partial \overline{V}}{\partial t}(t,x)+\frac{1}{2}\underset{%
\underline{\sigma }\leq \sigma \leq \overline{\sigma }}{\sup }\left\{ \sigma
^{2}x^{2}\frac{\partial ^{2}\overline{V}}{\partial x^{2}}(t,x)\right\}
+r_{t}x\frac{\partial \overline{V}}{\partial x}(t,x)-r_{t}\overline{V}(t,x)
&=&0,  \label{v.3.6} \\
\overline{V}(T,x) &=&\Phi (x).  \notag
\end{eqnarray}%
Then by the comparison theorem of PDEs, $\overline{V}(t,x)$ is the minimal
upper price outperforming all $\mu _{t}$ varying in $[\underline{\mu },%
\overline{\mu }]$ and $\sigma _{t}$ varying in $\left[ \underline{\sigma },%
\overline{\sigma }\right] $. There is no novelty in equation \eqref{v.3.6}
which is the so called Black-Scholes-Barenblatt (BSB) equation (\textrm{\cite%
{bar,alp}}). What is new is that, although we put risk preference and
uncertainty into stock appreciation $\mu $, the BSB equation does not
involve any variables that are affected by the risk preference of investors.
$\mu $ depends on risk preference and interval $[\underline{\mu },\overline{%
\mu }]$ determines mean-uncertainty. The higher the level risk and ambiguity
aversion by investors, the higher $\mu $ and the larger the uncertain
interval will be for any given stock. It is fortunate that $\mu $ happens to
drop out in the differential equation. So the risk preference of investors
and mean-uncertainty do not pose effects on our superhedging strategy. Thus
it is possible to consider risk-neutral \& mean-certain valuation under
model uncertainty.

\begin{remark}
Suppose that function $\Phi $ is a bounded continuous function. Assume that $%
\underline{\sigma }>0$. By \textrm{\cite{kry}} Theorem 6.4.3 or \textrm{\cite%
{wang}}, equation \eqref{v.3.6} has a $C^{1+\frac{\alpha }{2},2+\alpha }${}$%
([0$,{}${}${}$\,T)\times \mathbf{R})$-solution $u(t,x)$ for some $\alpha \in
(0,1)$. The uniqueness can be obtained from \textrm{\cite{ish}}. See also
\textrm{\cite{var}} for smooth solutions with locally Lipschitz terminal
condition.
\end{remark}

\begin{remark}
If there is uncertainty for the riskless interest rate, i.e., $r\in \lbrack
\underline{r},\overline{r}]$, then the superhedging PDE should be%
\begin{equation}
\frac{\partial \overline{V}}{\partial t}(t,x)+\frac{1}{2}\underset{\left(
r,\sigma \right) \in \lbrack \underline{r},\overline{r}]\times \left[
\underline{\sigma },\overline{\sigma }\right] }{\sup }\left\{ \sigma
^{2}x^{2}\frac{\partial ^{2}\overline{V}}{\partial x^{2}}(t,x)+r\left( x%
\frac{\partial \overline{V}}{\partial x}(t,x)-\overline{V}(t,x)\right)
\right\} =0.
\end{equation}
\end{remark}

\subsection{Discrete-path-dependent payoffs}

\label{v.sect3.2}We now consider the case of discrete-path-dependent payoff $%
\Phi (X_{t_{1}},\ldots ,X_{t_{n}})$ with $\Phi :\mathbf{R}^{N}\mapsto
\mathbf{R}$ for a partition $0=t_{0}<t_{1}<\cdots <t_{N}=T$. A typical
example is the arithmetic-mean Asian call $\Phi (X^{\left( N\right)
})=\left( \frac{1}{N}\sum_{i=0}^{N}X_{t_{i}}-K\right) ^{+}$, $K$ is the
fixed strike price. See \textrm{\cite{shreve}} for other kinds of
path-dependent options, such as lookback option, barrier option. For any $%
\mathbf{x:=}\left( x_{1},\ldots ,x_{N}\right) \in \mathbf{R}^{N}$ and $%
k=1,\ldots ,N$, we use the following notations:%
\begin{equation*}
\mathbf{x}^{\left( k\right) }\mathbf{:=}\left( x_{1},\ldots ,x_{k}\right)
\text{, }\mathbf{X}_{t}^{\left( k\right) }\mathbf{:=}\left( X_{t_{1}\wedge
t},\ldots ,X_{t_{k}\wedge t}\right) \text{, }\mathbf{X}^{\left( k\right) }%
\mathbf{:=}\left( X_{t_{1}},\ldots ,X_{t_{k}}\right) \text{.}
\end{equation*}

\bigskip Let $X_{t}$ denote the following stock price
\begin{equation}
dX_{t}=X_{t}\left( \mu _{t}dt+\sigma _{t}dW_{t}\right) ,  \label{v.3.8}
\end{equation}%
where $W$ is the standard Brownian motion under a given linear probability
space $(\Omega ,\mathcal{F},P)$; $\mu _{t},\sigma _{t}:$ $\mathbf{[0,+\infty
)}\mapsto \mathbf{R}$ valued in $[\underline{\mu },\overline{\mu }]$ and $%
\left[ \underline{\sigma },\overline{\sigma }\right] $ respectively. We now
derive the superhedging PDE on each\textrm{\ }$\left[ t_{k-1},t_{k}\right] $%
. Let $V^{k}(t,\mathbf{X}^{\left( k-1\right) },X_{t})_{t\in \left[
t_{k-1},t_{k}\right] }$ be the price of the option with payoff $\Phi (%
\mathbf{X}^{\left( N\right) })$. Assume also that $V^{k}\left( \cdot ,%
\mathbf{x}^{\left( k-1\right) },\cdot \right) \in C^{1,2}(\left[
t_{k-1},t_{k}\right] \times \mathbf{R})$. By It\^{o}'s formula, we have, $%
\forall t\in \left[ t_{k-1},t_{k}\right] $,%
\begin{eqnarray*}
dV^{k}(t,\mathbf{x}^{\left( k-1\right) },X_{t}) &=&\frac{\partial V^{k}}{%
\partial t}(t,\mathbf{x}^{\left( k-1\right) },X_{t})dt+X_{t}\frac{\partial
V^{k}}{\partial x}(t,\mathbf{x}^{\left( k-1\right) },X_{t})\left( \mu
_{t}dt+\sigma _{t}dW_{t}\right) \\
&&+\frac{1}{2}\sigma _{t}^{{\normalsize 2}}X_{t}^{2}\frac{\partial ^{2}V^{k}%
}{\partial x^{2}}(t,\mathbf{x}^{\left( k-1\right) },X_{t})dt.
\end{eqnarray*}%
By properties of integrals $dt$ and $dW_{t}$, we can replace $\mathbf{x}%
^{\left( k-1\right) }$ by $\mathbf{X}^{\left( k-1\right) }$ and get%
\begin{eqnarray*}
dV^{k}(t,\mathbf{X}^{\left( k-1\right) },X_{t}) &=&\frac{\partial V^{k}}{%
\partial t}(t,\mathbf{X}^{\left( k-1\right) },X_{t})dt+X_{t}\frac{\partial
V^{k}}{\partial x}(t,\mathbf{X}^{\left( k-1\right) },X_{t})\left( \mu
_{t}dt+\sigma _{t}dW_{t}\right) \\
&&+\frac{1}{2}\sigma _{t}^{{\normalsize 2}}X_{t}^{2}\frac{\partial ^{2}V^{k}%
}{\partial x^{2}}(t,\mathbf{X}^{\left( k-1\right) },X_{t})dt.
\end{eqnarray*}%
By an analogous procedure as in Section \ref{v.sect3.1}, the superhedging
price should satisfy%
\begin{eqnarray}
\frac{\partial V^{k}}{\partial t}(t,\mathbf{x}^{\left( k-1\right) },x)+ &&%
\frac{1}{2}\underset{\underline{\sigma }\leq \sigma \leq \overline{\sigma }}{%
\sup }\left\{ \sigma ^{2}x^{2}\frac{\partial ^{2}V^{k}}{\partial x^{2}}(t,%
\mathbf{x}^{\left( k-1\right) },x)\right\}  \label{v.3.10} \\
&+&r_{t}x\frac{\partial V^{k}}{\partial x}(t,\mathbf{x}^{\left( k-1\right)
},x)-r_{t}V^{k}(t,\mathbf{x}^{\left( k-1\right) },x)=0  \notag
\end{eqnarray}%
The sequence of PDEs $V^{k}$, $k=1,\ldots ,N$, is defined recursively in a
backward manner. The terminal conditions are defined respectively by

\begin{eqnarray}
V^{N}(T,\mathbf{x}^{\left( N-1\right) },x) &=&\Phi (\mathbf{x}^{\left(
N-1\right) },x),  \notag \\
&&\vdots  \notag \\
V^{k}(t_{k},\mathbf{x}^{\left( k-1\right) },x) &=&V^{k+1}(t_{k},\mathbf{x}%
^{\left( k-1\right) },x,x).  \label{v.3.11}
\end{eqnarray}%
As we see, the stock appreciation $\mu $ does not appear in \eqref{v.3.10}
due to delta-hedging. \eqref{v.3.10} can be used to super-hedge
discrete-path-dependent options. The existence and uniqueness of smooth
solutions for \eqref{v.3.10} and \eqref{v.3.11} can be guaranteed by \textrm{%
\cite{kry}} and \textrm{\cite{var}}. The randomness of $\mu _{t}$ and $%
\sigma _{t}$ does not influence PDE \eqref{v.3.10}.

\begin{remark}
\eqref{v.3.10} has another form:%
\begin{eqnarray*}
&&\frac{\partial V}{\partial t}(t,\mathbf{x}^{\left( k-1\right) },x)+\frac{1%
}{2}\sigma _{\max }^{2}\left( \frac{\partial ^{2}V}{\partial x^{2}}(t,%
\mathbf{x}^{\left( k-1\right) },x)\right) x^{2}\frac{\partial ^{2}V}{%
\partial x^{2}}(t,\mathbf{x}^{\left( k-1\right) },x) \\
&&\text{ \ \ }+r_{t}x\frac{\partial V^{k}}{\partial x}(t,\mathbf{x}^{\left(
k-1\right) },x)-r_{t}V(t,\mathbf{x}^{\left( k-1\right) },x)=0.
\end{eqnarray*}%
where $\sigma _{\max }^{2}\left( \frac{\partial ^{2}V}{\partial x^{2}}(t,%
\mathbf{x}^{\left( k-1\right) },x)\right) =\overline{\sigma }$, if $\frac{%
\partial ^{2}V}{\partial x^{2}}(t,\mathbf{x}^{\left( k-1\right) },x)\geq 0$;
$\sigma _{\max }^{2}\left( \frac{\partial ^{2}V}{\partial x^{2}}(t,\mathbf{x}%
^{\left( k-1\right) },x)\right) =\underline{\sigma }$, if $\frac{\partial
^{2}V}{\partial x^{2}}(t,\mathbf{x}^{\left( k-1\right) },x)<0$. As we see
above, the form of $\sigma _{t}$ does not pose effect on \eqref{v.3.10}.
Hence for the uncertain volatility model, we could just stand on the bounds
of the interval $\left[ \underline{\sigma },\overline{\sigma }\right] $.
\end{remark}

\subsection{General payoffs}

\label{v.sect3.3}From section \ref{v.sect3.2}, for discrete-path-dependent
payoffs, we can consider risk-neutral valuation. consider a stock price
process whose differential is
\begin{equation}
dX_{t}=X_{t}\left( r_{t}dt+dB_{t}\right) ,0\leq t\leq T,  \label{v.3.111}
\end{equation}%
where $\left( B_{t}\right) $ a $G_{[\underline{\sigma }^{2},\overline{\sigma
}^{2}]}$-Brownian motion, $r_{t}$ is the interest rate from $\left[ 0\mathbf{%
,}T\right] $ to $\mathbf{R}$. For the solution $V^{k}\left( \cdot ,\mathbf{x}%
^{\left( k-1\right) },\cdot \right) \in C^{1,2}(\left[ t_{k-1},t_{k}\right]
\times \mathbf{R})$ of \eqref{v.3.10}, applying It\^{o}'s formula to $%
V^{k}\left( t,\mathbf{x}^{\left( k-1\right) },X_{t}\right) $, and
substituting $\mathbf{x}^{\left( k-1\right) }$ by $\mathbf{X}^{\left(
k-1\right) }$ we deduce that for $\left[ t_{k-1},t_{k}\right] $,%
\begin{eqnarray*}
Y_{t}^{k} &=&V^{k}\left( t,\mathbf{X}_{t}^{\left( k\right) }\right) \text{, }
\\
Z_{t}^{k} &=&X_{t}\frac{\partial V^{k}}{\partial x}\left( t,\mathbf{X}%
_{t}^{\left( k\right) }\right) \text{, } \\
K_{t}^{k} &=&\frac{1}{2}\int_{0}^{t}\underset{\underline{\sigma }\leq \sigma
\leq \overline{\sigma }}{\sup }\left\{ \sigma ^{2}X_{t}^{2}\frac{\partial
^{2}V^{k}}{\partial x^{2}}\left( t,\mathbf{X}_{t}^{\left( k\right) }\right)
\right\} dt-\frac{1}{2}\int_{0}^{t}X_{t}^{2}\frac{\partial ^{2}V^{k}}{%
\partial x^{2}}\left( t,\mathbf{X}_{t}^{\left( k\right) }\right)
d\left\langle B\right\rangle _{t}
\end{eqnarray*}%
satisfy the following risk-neutral BSDE:%
\begin{equation*}
Y_{t}=Y_{t_{k}}-\int_{t}^{t_{k}}{\normalsize r_{s}}Y_{s}ds-%
\int_{t}^{t_{k}}Z_{s}dB_{s}+\int_{t}^{t_{k}}dK_{s},\ t\in \left[
t_{k-1},t_{k}\right] ,
\end{equation*}%
which coincides with the following BSDE:%
\begin{equation*}
Y_{t}=\Phi (\mathbf{X}^{\left( N\right) })-\int_{t}^{T}{\normalsize r_{s}}%
Y_{s}ds-\int_{t}^{T}Z_{s}dB_{s}+\int_{t}^{T}dK_{s},\ t\in \left[ 0,T\right]
\end{equation*}%
on $\left[ t_{k-1},t_{k}\right] $. For a general payoff $\xi \in L{_{G}^{%
\mathrm{1}}}(\Omega _{T})$, it can be approximated by a sequence of $\Phi
^{n}(\mathbf{X}^{\left( n\right) })$, $n=1,2,\ldots $, for appropriate
diffusion process $\left( X_{t}\right) $. We can check that the following
sequence%
\begin{equation*}
Y_{t}^{n}=\Phi ^{n}(\mathbf{X}^{\left( n\right) })-\int_{t}^{T}{\normalsize %
r_{s}}Y_{s}^{n}ds-\int_{t}^{T}Z_{s}^{n}dB_{s}+\int_{t}^{T}dK_{s}^{n},\ t\in %
\left[ 0,T\right]
\end{equation*}%
converges to%
\begin{equation}
Y_{t}=\xi -\int_{t}^{T}{\normalsize r_{s}}Y_{s}ds-\int_{t}^{T}Z_{s}dB_{s}+%
\int_{t}^{T}dK_{s},\ t\in \left[ 0,T\right]  \label{v.3.12}
\end{equation}%
in the space $\mathcal{M}{_{G}^{\mathrm{2}}}(0,{\mathit{T}})$. So for a
general payoff, its risk-neutral superhedging price exists and can be
calculated by%
\begin{equation*}
V_{t}=D_{t}^{-1}\mathbb{E}\left[ D_{T}\xi |\mathcal{F}_{t}\right]
\end{equation*}%
where $D_{t}={\normalsize \exp }\left\{ -{\normalsize \int_{0}^{t}r_{s}ds}%
\right\} $.

\begin{remark}
We can replace $r_{t}$ in \eqref{v.3.111} by the random interest rate $r(%
\mathbf{X}_{t}^{\left( k\right) })$, then by approximation, $\left(
r_{t}\right) $ in \eqref{v.3.12} can be an adapted stochastic process.
\end{remark}

\section{Superhedging and subhedging under volatility uncertainty and
arbitrage ambiguity}

\label{v.sect4}Now we consider the hedging problem by Peng's G-stochastic
analysis in the risk-neutral \& mean-certain world. Let $\mathcal{P}=\left\{
P\right\} $ be the set of risk-neutral measures and $\mathbb{E}$ the
corresponding risk-neutral sublinear expectation. Let $(B_{t})$ denote the $%
G_{[\underline{\sigma }^{2},\overline{\sigma }^{2}]}$-Brownian motion under $%
\mathbb{E}$. Let ${\mathcal{F}}_{t}$ be the minimal $\sigma $-algebra $\cap
_{r>t}\sigma \left\{ B_{s},s\leq r\right\} $. Here $T$ is a fixed time. We
consider a financial market with two assets. One of them is a locally
riskless asset (the bank account) with price per unit $(C_{t})$ governed by
the equation%
\begin{equation}
dC_{t}=C_{t}r_{t}dt,
\end{equation}%
where $\left( r_{t}\right) $ is the nonnegative short rate. In addition to
the bank account, consider a stock price process whose differential is
\begin{equation}
dS_{t}=S_{t}\left( r_{t}dt+dB_{t}\right) ,0\leq t\leq T,  \label{v.4.2}
\end{equation}%
where $\left( B_{t}\right) $ a $G_{[\underline{\sigma }^{2},\overline{\sigma
}^{2}]}$-Brownian motion. Stochastic process $\left( r_{t}\right) $ is
allowed to be bounded ${\mathcal{F}}_{t}$-progressively measurable process
in $\mathcal{M}{_{G}^{\mathrm{2}}}(0,{\mathit{T}})$.\footnote{$\mathcal{M}{%
_{G}^{\mathrm{2}}}$ is the space consisting of square-integrable random
variables such that the G-stochastic integral is well defined. See Appendix
A for details.}

The market can not be complete because of the uncertain volatility. So
investors could not expect to replicate exactly any general contingent claim
and have to choose some criterion to hedge the claim.

\subsection{Superhedging for the option seller}

\label{v.sect4.1}Let $\xi $ be an ${\mathcal{F}}_{T}$-measurable random
variable which represents the payoff at time T of a derivative security. We
allow this payoff to be path-dependant, i.e., to depend on anything that
occurs between times $0$ and $T$. We now give the definition of
superstrategy under model uncertainty.

\begin{definition}
\label{v.def4.1}A K-financing superstrategy against a contingent claim $\xi $
under model uncertainty is a vector process ($V,\pi ,K$), where $V$ is the
managing price, $\pi $ is the portfolio process, and $K$ is the pricing
error, such that%
\begin{align}
dV_{t}& =r_{t}V_{t}dt+\pi _{t}dB_{t}-dK_{t},\ q.s.\   \label{v.4.4} \\
V_{T}& =\xi ,\ and\ \int_{0}^{T}\left\vert \pi _{t}\right\vert ^{2}dt<\infty
,q.s.  \notag
\end{align}%
where $K$ is an increasing, right-continuous adapted process q.s. with $%
K_{0}=0$.
\end{definition}

\begin{remark}
Any superstrategy defined by Definition \ref{v.def4.1} satisfies%
\begin{equation*}
V_{t}\geq v_{t}^{P},\forall t\in \lbrack 0,T],\forall P\in \mathcal{P},P-a.s.
\end{equation*}%
due to the comparison of BSDEs (\textrm{\cite{epq}}), where $\left(
v_{t}^{P}\right) $ solves%
\begin{equation}
dv_{t}^{P}=r_{t}v_{t}^{P}dt+\pi _{t}^{P}\sigma _{t}^{P}dW_{t}^{P},\
v_{T}^{P}=\xi ,\ P-a.s.  \label{v.4.5}
\end{equation}%
with $\sigma _{t}^{P}$ being $\mathcal{F}_{t}^{P}$-adapted process valued in
$[\underline{\sigma },\overline{\sigma }]$ and $\left( W_{t}^{P}\right) $
the standard Brownian motion in a linear expectation space $(\Omega ,(%
\mathcal{F}_{t}^{P})_{t\geq 0},{\normalsize E}^{P})$.
\end{remark}

\begin{definition}
\label{v.def4.2}There is an arbitrage for a superstrategy ($V_{t},\pi
_{t},K_{t}$) if the value process $\left( V_{t}\right) $ satisfies $V_{0}=0$
and
\begin{equation}
V_{T}\geq 0,q.s.\text{\ and\ }P[V_{T}>0]>0\text{,\ for\ at\ least\ one\ }%
P\in \mathcal{P}.  \label{v.4.6}
\end{equation}
\end{definition}

\begin{theorem}
\label{v.th4.1}The solution triple ($V_{t},\pi _{t},K_{t}$) to BSDE %
\eqref{v.4.4} is the minimal superstrategy with no-arbitrage. The
\textquotedblleft minimal\textquotedblright\ means that for any other
superstrategy ($V_{t}^{\prime },\pi _{t}^{\prime },K_{t}^{\prime }$), we
have $V_{t}\leq V_{t}^{\prime },\forall t,q.s.$.
\end{theorem}

\textsc{Proof}: \ Let ($V_{t},\pi _{t},K_{t}$) be the unique triple
satisfying the BSDE \eqref{v.4.4} and $V_{T}=\xi $ with $(-K_{t})$ being a
continuously nonincreasing G-martingale. Obviously ($V_{t},\pi _{t},K_{t}$)
is a superstrategy according Definition \ref{v.def4.1}. Furthermore, by
Theorem \ref{v.thB.1}, we have
\begin{equation}
V_{t}=D_{t}^{-1}\mathbb{E}\left[ D_{T}\xi |\mathcal{F}_{t}\right]
\label{v.4.7}
\end{equation}%
where $D_{t}={\normalsize \exp }\left\{ -{\normalsize \int_{0}^{t}r_{s}ds}%
\right\} $.

Let ($V_{t}^{\prime },\pi _{t}^{\prime },K_{t}^{\prime }$) be another
superstrategy defined by Definition \ref{v.def4.1} with ($K_{t}^{\prime }$)
being an increasing, right-continuous adapted process q.s. and $%
K_{0}^{\prime }=0$. Applying It\^{o}'s formula to $D_{t}V_{t}^{\prime }$, we
obtain that
\begin{align*}
d(D_{t}V_{t}^{\prime })& =D_{t}\left[ r_{t}V_{t}^{\prime }dt+\pi
_{t}^{\prime }dB_{t}-dK_{t}^{\prime }\right] -V_{t}^{\prime }D_{t}r_{t}dt \\
& =D_{t}\pi _{t}^{\prime }dB_{t}-D_{t}dK_{t}^{\prime }.
\end{align*}%
Note that $(D_{t}V_{t}^{\prime }+\int_{0}^{t}D_{s}dK_{s}^{\prime })$ is a
G-martingale. Therefore
\begin{align}
V_{t}^{\prime }& =D_{t}^{-1}\left( \mathbb{E}\left[ D_{T}\xi
+\int_{0}^{T}D_{s}dK_{s}^{\prime }|\mathcal{F}_{t}\right] -%
\int_{0}^{t}D_{s}dK_{s}^{\prime }\right)  \notag \\
& =D_{t}^{-1}\left( \mathbb{E}\left[ D_{T}\xi
+\int_{t}^{T}D_{s}dK_{s}^{\prime }|\mathcal{F}_{t}\right] \right) .
\label{v.4.8}
\end{align}%
Since $\int_{t}^{T}D_{t,s}dK_{s}^{\prime }\geq 0,\forall t\in \lbrack
0,T],q.s.$, then by the monotonicity of conditional G-expectation, we obtain%
\begin{equation*}
V_{t}^{\prime }\geq V_{t},\forall t\in \lbrack 0,T],q.s.
\end{equation*}%
Hence ($V_{t},\pi _{t},K_{t}$) is the minimal superstrategy covering every
probabilistic model.

If the terminal position
\begin{equation*}
V_{T}\geq 0,q.s.\text{\ and\ }\exists P\in \mathcal{P}\text{,\ such\ that\ }%
P[V_{T}>0]>0,
\end{equation*}%
then
\begin{equation*}
V_{0}=\mathbb{E}\left[ D_{T}V_{T}\right] =\underset{P\in \mathcal{P}}{\sup }%
{\normalsize E}_{P}\left[ D_{T}V_{T}\right] >0.
\end{equation*}%
So the superstrategy ($V,\pi ,K$) is arbitrage-free. $\Box $

By Theorem \ref{v.th4.1}, we know that a hedging strategy is a minimal
superstrategy under model uncertainty if and only if $\left( -K_{t}\right) $
is a G-martingale with finite variance such that \eqref{v.4.4} holds and $%
V_{T}=\xi $. We will give more explicit explanation for $K$ in the language
of P\&L, see Section \ref{v.sect5.3}.

\begin{remark}
(K-financing and self-financing) The solution triple ($V_{t},\pi _{t},K_{t}$%
) to BSDE \eqref{v.4.4} is a K-financing superstrategy with $(-K_{t})$ being
a continuously nonincreasing G-martingale. Clearly ($V_{t},\pi _{t},K_{t}$)
is not necessary a self-financing strategy because the cumulative
consumption $(K_{t})$ is a nonnegative increasing process q.s. with $K_{0}=0$%
. Since $(-K_{t})$ is a G-martingale, there exists a probability measure $P$
such that
\begin{equation*}
0=K_{0}=\mathbb{E}\left[ -K_{T}\right] ={\normalsize E}_{P}\left[ -K_{T}%
\right] .
\end{equation*}%
Thus $K_{T}\equiv 0$ and $K_{t}\equiv 0$, $P$-$a.s.$, for each $t$. So a
K-financing superstrategy ($V_{t},\pi _{t},K_{t}$) is a self-financing
strategy under some $P\in \mathcal{P}$. Certainly if any other $P^{\prime
}\in \mathcal{P}$ is equivalent to $P$, then $K_{t}\equiv 0$, $\mathcal{P}$-$%
q.s.$, for each $t$ and ($V_{t},\pi _{t}$) is a self-financing strategy
under each $P\in \mathcal{P}$. So for a set probabilities $\mathcal{P}$
which consists of mutually singular probability measures, in general, we can
not find a universal self-financing hedging strategy, which leads to the
incompleteness of a financial market.
\end{remark}

\subsection{Subhedging for the option buyer}

\label{v.sect4.2}Usually an option buyer puts more attention on
substrategies, in particular the maximal substrategy which can be viewed as
the maximal amount that the buyer of the option is willing to pay at time 0
such that he/she is sure to cover at time $T$, the debt he/she incurred at
time 0.

\begin{definition}
\label{v.def4.4}A K-financing substrategy against a contingent claim $\xi $
under model uncertainty is a vector process ($\widetilde{V},\widetilde{\pi },%
\widetilde{K}$), where $\widetilde{V}$ is the market value, $\widetilde{\pi }
$ is the portfolio process, and $\widetilde{K}$ is the pricing error, such
that%
\begin{equation}
d\widetilde{V}_{t}=r_{t}\widetilde{V}_{t}dt+\widetilde{\pi }_{t}dB_{t}+d%
\widetilde{K}_{t}\text{,\ q.s.\ and\ }\widetilde{V}_{T}=\xi ,\
\int_{0}^{T}\left\vert \widetilde{\pi }_{t}\right\vert ^{2}dt<\infty \text{,
q.s.}  \label{v.4.9}
\end{equation}%
where $\widetilde{K}$ is an increasing, right-continuous ${\mathcal{F}}_{t}$%
-progressively measurable process q.s. with $\widetilde{K}_{0}=0$.
\end{definition}

\begin{remark}
Any substrategy defined by Definition \ref{v.def4.4} satisfies%
\begin{equation*}
\widetilde{V}_{t}\leq v_{t}^{P},\forall t\in \lbrack 0,T],\forall P\in
\mathcal{P},P\text{-a.s.}
\end{equation*}%
where $\left( v_{t}^{P}\right) $ solves BSDE \eqref{v.4.5}.
\end{remark}

Let $\widetilde{\mathbb{E}}\left[ \cdot|\mathcal{F}_{t}\right] :=-\mathbb{E}%
\left[ -\cdot|\mathcal{F}_{t}\right] $. Then one can easily check that $%
\widetilde{\mathbb{E}}$ satisfies the following super-additivity:
\begin{equation*}
\widetilde{\mathbb{E}}\mathbb{[}X+Y|\mathcal{F}_{t}\mathbb{]}\geq \widetilde{%
\mathbb{E}}\mathbb{[}X|\mathcal{F}_{t}\mathbb{]}+\widetilde {\mathbb{E}}%
\mathbb{[}Y|\mathcal{F}_{t}\mathbb{]}
\end{equation*}
and shares all other properties of $\mathbb{E}$.

\begin{theorem}
\label{v.th4.3}The maximal substrategy ($\widetilde{V},\widetilde{\pi },%
\widetilde{K}$) satisfying
\begin{equation}
d\widetilde{V}_{t}=r_{t}\widetilde{V}_{t}dt+\widetilde{\pi }_{t}dB_{t}+d%
\widetilde{K}_{t},\ \widetilde{V}_{T}=\xi  \label{v.4.10}
\end{equation}%
where $(\widetilde{K}_{t})$ is a continuous, increasing process with $%
\widetilde{K}_{0}=0$ and $(\widetilde{K}_{t})$ being a martingale under $%
\widetilde{\mathbb{E}}$. More explicitly we have for any $t\in \lbrack 0,T]$%
,
\begin{equation*}
\widetilde{V}_{t}=D_{t}^{-1}\widetilde{\mathbb{E}}\left[ D_{T}\xi |\mathcal{F%
}_{t}\right] ,\ q.s.
\end{equation*}%
The \textquotedblleft maximal\textquotedblright\ means that for any other
substrategy ($\widetilde{V}_{t}^{\prime },\widetilde{\pi }_{t}^{\prime },%
\widetilde{K}_{t}^{\prime }$), we have $\widetilde{V}_{t}\geq \widetilde{V}%
_{t}^{\prime },\forall t,q.s.$.
\end{theorem}

\textsc{Proof}: \ Let ($\widetilde{V},\widetilde{\pi },\widetilde{K}$) be
the unique triple satisfying BSDE \eqref{v.4.10} with $(\widetilde{K}_{t})$
being a continuous, increasing martingale under $\widetilde{\mathbb{E}}\left[
\cdot |\mathcal{F}_{t}\right] $. Obviously ($\widetilde{V},\widetilde{\pi },%
\widetilde{K}$) is a substrategy according Definition \ref{v.def4.4}.
Applying It\^{o}'s formula to $D_{t}\widetilde{V}_{t}$, we get that%
\begin{equation*}
\widetilde{V}_{t}=D_{t}^{-1}\widetilde{\mathbb{E}}\left[ D_{T}\xi |\mathcal{F%
}_{t}\right] \text{,\ \textit{q.s}.}
\end{equation*}

Let ($\widetilde{V}^{\prime },\widetilde{\pi }^{\prime },\widetilde{K}%
^{\prime }$) be another substrategy defined by Definition \ref{v.def4.4}
with $\widetilde{K}^{\prime }$ being an increasing, right-continuous adapted
process q.s. and $\widetilde{K}_{0}^{\prime }=0$. By direct calculation
similarly to equation \eqref{v.4.8}, we get
\begin{equation*}
\widetilde{V}_{t}^{\prime }=D_{t}^{-1}\widetilde{\mathbb{E}}\left[ D_{T}\xi
-\int_{t}^{T}D_{s}d\widetilde{K}_{s}^{\prime }|\mathcal{F}_{t}\right] .
\end{equation*}%
Since $-\int_{t}^{T}D_{s}dK_{s}^{\prime }\leq 0,\forall t\in \lbrack
0,T],q.s.$, then by the monotonicity of conditional expectation $\widetilde{%
\mathbb{E}}\left[ \cdot |\mathcal{F}_{t}\right] $, we obtain%
\begin{equation*}
\widetilde{V}_{t}^{\prime }\leq \widetilde{V}_{t},\forall t\in \lbrack
0,T],q.s.
\end{equation*}%
Therefore ($\widetilde{V},\widetilde{\pi },\widetilde{K}$) is the maximal
substrategy under every probabilistic model. $\Box $

\begin{remark}
\label{v.rem4.5}For a substrategy ($\widetilde{V},\widetilde{\pi },%
\widetilde{K}$) satisfying \eqref{v.4.10}, condition \eqref{v.4.6} does not
guarantee no-arbitrage. Even condition \eqref{v.4.6} of arbitrage is
replaced by
\begin{equation}
\ \widetilde{V}_{T}\geq 0\text{, q.s.\ and\ for\ \textbf{all}\ }P\in
\mathcal{P},\ P\mathcal{[}\widetilde{V}_{T}>0]>0\text{,}  \label{v.4.11}
\end{equation}%
then still there may be an arbitrage opportunity. In fact if \eqref{v.4.11}
holds, we have $\forall \ P\in \mathcal{P}$, ${\normalsize E}_{P}\left[
D_{T}\xi \right] >0$. But after taking infimum, perhaps%
\begin{equation*}
\widetilde{V}_{0}=\underset{P\in \mathcal{P}}{\inf }{\normalsize E}_{P}\left[
D_{T}\xi \right] =0.
\end{equation*}
\end{remark}

So we have to redefine the notion of arbitrage for sub-hedging strategies.

\begin{definition}
\label{v.def4.50}There is an arbitrage for a substrategy ($\widetilde{V},%
\widetilde{\pi },\widetilde{K}$) satisfying \eqref{v.4.10}, if the value
process $(\widetilde{V}_{t})$ satisfies $\widetilde{V}_{0}=0$ and
\begin{equation}
\widetilde{V}_{T}\geq 0\text{, q.s.\ and\ }\underset{P\in \mathcal{P}}{\inf }%
P[\widetilde{V}_{T}>0]>0.  \label{v.4.12}
\end{equation}
\end{definition}

Under the above definition, we have,

\begin{theorem}
\label{v.th4.40}The substrategy ($\widetilde{V},\widetilde{\pi },\widetilde{K%
}$) is arbitrage-free.
\end{theorem}

\textsc{Proof}: \ If \eqref{v.4.12} holds, then by the strict comparison
theorem in \textrm{\cite{li}}\footnote{%
The strict comparison theorem\textrm{\ }says that: for $\xi ^{1},\xi ^{2}\in
L_{G}^{1}\left( \Omega \right) $, if $\xi ^{1}\geq \xi ^{2}$ and $\underset{%
P\in \mathcal{P}}{\inf }P[\xi ^{1}>\xi ^{2}]>0$, then $\mathbb{E}\left[ \xi
^{1}\right] >\mathbb{E}\left[ \xi ^{2}\right] $ and $\widetilde{\mathbb{E}}%
\left[ \xi ^{1}\right] >\widetilde{\mathbb{E}}\left[ \xi ^{2}\right] $.}, we
have $\widetilde{V}_{0}=\widetilde{\mathbb{E}}\left[ D_{T}\xi \right] =%
\underset{P\in \mathcal{P}}{\inf }{\normalsize E}_{P}\left[ D_{T}\xi \right]
>0$. Thus there is no arbitrage for the substrategy. $\Box $

\subsection{Put-call parity}

\label{v.sect4.4}In a complete financial market, there is a parity relation
between a pair of European call option and European put option underlying
the same stock $S$ and with the same expiration date and strike price. We
now consider similar parity relation for superhedging strategies in an
incomplete market. The superhedging prices of a European call option and a
European put option underlying the same stock $S$ and sharing the same
strike price $L$ are given by
\begin{equation*}
c_{t}=(S_{T}-L)^{+}-\int_{t}^{T}{\normalsize r_{s}}c_{s}ds-\int_{t}^{T}\pi
_{s}^{c}dB_{s}+\int_{t}^{T}dK_{s}^{c},\ t\in \lbrack 0,T],
\end{equation*}%
and
\begin{equation*}
p_{t}=(L-S_{T})^{+}-\int_{t}^{T}{\normalsize r_{s}}p_{s}ds-\int_{t}^{T}\pi
_{s}^{p}dB_{s}+\int_{t}^{T}dK_{s}^{p},\ t\in \lbrack 0,T],
\end{equation*}%
where $L\in \mathbf{R}^{+}$ is the strike price and $\left( S_{t}\right) $
is the stock price following%
\begin{equation*}
dS_{t}=S_{t}\left( r_{t}dt+dB_{t}\right) ,t\geq 0,
\end{equation*}%
where $r_{t}$ is $\mathcal{F}_{t}$-measurable bounded processes belonging to
${\mathcal{M}_{G}^{\mathrm{2}}}$.

\begin{theorem}
\label{v.th4.4}Let $c_{t}$ and $p_{t}$ be the superhedging prices of a
European call option and a European put option underlying the same stock $S$
and sharing the same strike price $L$. Then
\begin{equation*}
c_{t}+L\cdot \exp \left\{ -\int_{t}^{T}{\normalsize r_{s}}ds\right\}
=p_{t}+S_{t},\ q.s.
\end{equation*}%
Similarly the parity relation also holds for subhedging prices.
\end{theorem}

\textsc{Proof}: \ Let $L_{t}=L\cdot \exp \left\{ -\int_{t}^{T}{\normalsize %
r_{s}}ds\right\} $. Then
\begin{equation*}
L_{t}=L-\int_{t}^{T}{\normalsize r_{s}}L_{s}ds.
\end{equation*}%
By doing summation, we get%
\begin{equation*}
c_{t}+L_{t}=(S_{T}-L)^{+}+L-\int_{t}^{T}{\normalsize r_{s}}\left(
c_{s}+L_{s}\right) ds-\int_{t}^{T}\pi _{s}^{c}dB_{s}+\int_{t}^{T}dK_{s}^{c},
\end{equation*}%
and
\begin{equation*}
p_{t}+S_{t}=(L-S_{T})^{+}+S_{T}-\int_{t}^{T}{\normalsize r_{s}}\left(
p_{s}+S_{s}\right) ds-\int_{t}^{T}\left( \pi _{s}^{p}+S_{s}\right)
dB_{s}+\int_{t}^{T}dK_{s}^{p}.
\end{equation*}%
Observing that $(S_{T}-L)^{+}+L=(L-S_{T})^{+}+S_{T}=\max \left\{
L,S_{T}\right\} $ and the uniqueness of solution (See Theorem \ref{v.thB.1})
of the following BSDE%
\begin{equation*}
y_{t}=\max \left\{ L,S_{T}\right\} -\int_{t}^{T}{\normalsize r_{s}}%
y_{s}ds-\int_{t}^{T}{\normalsize z}_{s}dB_{s}+\int_{t}^{T}dK_{s},\ t\in
\lbrack 0,T],
\end{equation*}%
we deduce that the put-call parity
\begin{equation*}
c_{t}+L_{t}=p_{t}+S_{t}
\end{equation*}%
holds. $\Box $

\subsection{Asset with strictly non-zero upper price and generalized
geometric G-Brownian motion}

\label{v.sect4.5}

\begin{definition}
\label{v.def4.5}A sublinear expectation $\mathbb{E}$ is said to be
risk-neutral if the discounted stock price $\left( D_{t}S_{t}\right) $
(paying no dividend) is a symmetric G-martingale under $\mathbb{E}$.
\end{definition}

\begin{proposition}
\label{v.pr4.1}Let $\mathbb{E}$ be a risk-neutral sublinear expectation in a
market model. Then the upper price of every discounted portfolio is a
G-martingale (not necessarily symmetric) under $\mathbb{E}$.
\end{proposition}

\textsc{Proof}: \ Let $\left( B_{t}\right) $ be the G-Brownian motion under $%
\mathbb{E}$. Assume that the stock price follows $\frac{dS_{t}}{S_{t}}%
=r_{t}dt+dB_{t}$. Then the upper price of a portfolio follows%
\begin{eqnarray*}
dV_{t} &=&r_{t}(V_{t}-\pi _{t})dt+\pi _{t}\frac{dS_{t}}{S_{t}}-dK_{t} \\
&=&r_{t}V_{t}dt+\pi _{t}dB_{t}-dK_{t},
\end{eqnarray*}%
where $\left( -K_{t}\right) $ is a continuous nonincreasing G-martingale
under $\mathbb{E}$. Then the differential of the discounted upper price is
\begin{eqnarray*}
d\left( D_{t}V_{t}\right) &=&D_{t}dV_{t}+V_{t}dD_{t}=D_{t}\left[
r_{t}(V_{t}-\pi _{t})dt+\pi _{t}\frac{dS_{t}}{S_{t}}-dK_{t}\right]
+V_{t}dD_{t} \\
&=&D_{t}\left[ r_{t}(V_{t}-\pi _{t})dt+\pi _{t}\left( r_{t}dt+dB_{t}\right)
-dK_{t}\right] -r_{t}D_{t}V_{t}dt \\
&=&\frac{\pi _{t}}{S_{t}}d\left( D_{t}S_{t}\right) -D_{t}dK_{t}.
\end{eqnarray*}%
Under the risk-neutral sublinear expectation $\mathbb{E}$, $\left(
D_{t}S_{t}\right) $ is a symmetric G-martingale, $\left(
-\int_{0}^{t}D_{s}dK_{s}\right) $ is a G-martingale with finite variance.
Hence the process $\left( D_{t}V_{t}\right) $ must be a G-martingale. $\Box $

\begin{definition}
\label{v.def4.6}A process $\left( V_{t}\right) $ is called a geometric
G-Brownian motion if it follows%
\begin{equation}
dV_{t}=V_{t}\left( r_{t}dt+\alpha _{t}dK_{t}+\theta _{t}dB_{t}\right)
\end{equation}%
where $\left( B_{t}\right) $ is a $G$-Brownian motion, $\left( K_{t}\right) $
is a right-continuous increasing process, $r_{t}\in M{_{G}^{\mathrm{1}}}$ , $%
\alpha _{t}\in M{_{G}^{\mathrm{1}}}$ and $(\underset{0\leq t\leq T}{\sup }%
\left\vert \alpha _{t}\right\vert )\cdot K_{T}<\infty $, $\theta _{t}\in M{%
_{G}^{\mathrm{2}}}$. Or equivalently%
\begin{equation*}
V_{t}=V_{0}\exp \left\{ \int_{0}^{t}\theta
_{s}dB_{s}+\int_{0}^{t}r_{s}ds+\int_{0}^{t}\alpha _{s}dK_{s}-\frac{1}{2}%
\int_{0}^{t}\theta _{s}^{2}d\left\langle B\right\rangle _{s}\right\} .
\end{equation*}
\end{definition}

An asset with strictly non-zero upper price is a security paying $V_{T}$ at
time T whose upper price $V_{t}\neq 0,q.s.$ for each $t\in \lbrack 0,T]$.

\begin{theorem}
\label{v.th4.5}The upper price of an asset is strictly non-zero if and only
if the upper price is a generalized geometric G-Brownian motion with $%
V_{0}\neq 0$.
\end{theorem}

\textsc{Proof}: \ Let $\mathbb{E}$ be the unique risk-neutral sublinear
expectation. Then $\mathbb{E}\left[ D_{T}V_{T}|\mathcal{F}_{t}\right]
=D_{t}V_{t},q.s.$ for each $t\in \lbrack 0,T]$. By the Martingale
Representation Theorem, there exists an adapted process $\left( Z_{t}\right)
$ and nonincreasing $\mathbb{E}$-martingale $\left( -K_{t}\right) $ such that%
\begin{equation*}
D_{t}V_{t}=\mathbb{E}\left[ D_{T}V_{T}|\mathcal{F}_{t}\right]
=V_{0}+\int_{0}^{t}Z_{s}dB_{s}-K_{t},
\end{equation*}%
where $\left( B_{t}\right) $ is a G-Brownian motion under $\mathbb{E}$. Thus
the differential of $(V_{t})$ is%
\begin{equation*}
dV_{t}=r_{t}V_{t}dt+D_{t}^{-1}Z_{t}dB_{t}-D_{t}^{-1}dK_{t}
\end{equation*}%
Set $\theta _{t}=\frac{D_{t}^{-1}Z_{t}}{V_{t}}$, $\alpha _{t}=\frac{%
D_{t}^{-1}}{V_{t}}$. Then%
\begin{equation*}
dV_{t}=V_{t}\left( r_{t}dt+\alpha _{t}dK_{t}+\theta _{t}dB_{t}\right)
\end{equation*}%
Or%
\begin{equation*}
V_{t}=V_{0}\exp \left\{ \int_{0}^{t}\theta
_{s}dB_{s}+\int_{0}^{t}r_{s}ds+\int_{0}^{t}\alpha _{s}dK_{s}-\frac{1}{2}%
\int_{0}^{t}\theta _{s}^{2}d\left\langle B_{s}\right\rangle \right\} .
\end{equation*}

The sufficiency is obvious. $\Box $

\begin{corollary}
Every asset with strictly positive payoff is a generalized geometric
G-Brownian motion.
\end{corollary}

\textsc{Proof}: \ Since the payoff $V_{T}>0,q.s.$, by the risk-neutral
pricing formula, for each $t\in \lbrack 0,T]$,%
\begin{equation*}
V_{t}=D_{t}^{-1}\mathbb{E}_{\mathcal{Q}}\left[ D_{T}V_{T}|\mathcal{F}_{t}%
\right] >0,q.s.
\end{equation*}%
Then this corollary is obtained by Theorem \ref{v.th4.5}. $\Box $

\section{Results in Markovian setting}

\label{v.sect5}In this section, we consider some results using the
state-dependent BSB equation.

\subsection{Interpretation of $\protect\eta $ and $K$}

\label{v.sect5.3}Why do $K$ and $\eta $ arise when we super-hedge under
volatility uncertainty? Do they have certain sound financial meaning? We
have given a rough explanation of the finite-variance term $K$ in BSDE %
\eqref{v.4.4}. In Markovian setting, $K$ has a concrete decomposition: $%
K_{t}=\int_{0}^{t}\left[ 2G\left( \eta _{s}\right) ds-\eta _{s}d\left\langle
B\right\rangle _{s}\right] $, where $\eta _{t}=\frac{1}{2}S_{t}^{2}\Gamma
_{t}$, $\Gamma _{t}=\frac{\partial ^{2}u}{\partial S^{2}}(S_{t})$ is the
Gamma of the option with payoff $\Phi (S_{T})$. Obviously

$\bullet$ $\eta$ corresponds to Gamma $\Gamma$ of the option, while we have
known that $Z$ corresponds to Delta $\Delta$ of the option.

In the classical Black-Scholes-Merton model, when a trader uses the
Black-Scholes formula to sell and dynamically hedge a call option at
managing volatility $\sigma _{t}$, if the realized volatility is lower than
the managing volatility, the corresponding P\&L will be non negative. An
application of It\^{o} formula shows us that the instantaneous P\&L\footnote{%
See \textrm{\cite{mj}}, \textrm{\cite{js}} for the definition and derivation
of P\&L.} of being short a delta-hedged option reads%
\begin{equation}
P\&L_{(t,t+dt)}=\frac{1}{2}S_{t}^{2}\Gamma _{t}\left[ \sigma
_{t}^{2}dt-\left( \frac{dS_{t}}{S_{t}}\right) ^{2}\right]
\end{equation}%
where $\sigma _{t}$ is the managing volatility, i.e. the volatility at which
the option is sold and $\left( \frac{dS_{t}}{S_{t}}\right) ^{2}$ represents
the realized variance over the period $[t,t+dt]$. $\Gamma $ is positive for
a call option and an upper bound of the realized volatility is enough to
grant a profit (conversely, a lower bound for option buyers).

For an option with payoff $\Phi (S_{T})$ and with volatility fluctuating in
interval [$\underline{\sigma },\overline{\sigma }$] at each time $t$,
investors seek for a managing policy yielding a non negative P\&L whatever
the realized path. So investors sell the option at maximal volatility in
some sense such that the maximal instantaneous P\&L of being short a
delta-hedged option should be in form of
\begin{equation}
P\&L_{(t,t+dt)}=\frac{1}{2}\underset{\underline{\sigma }\leq \sigma _{t}\leq
\overline{\sigma }}{\sup }\{\sigma _{t}^{2}S_{t}^{2}\Gamma _{t}\}dt-\frac{1}{%
2}S_{t}^{2}\Gamma _{t}d\left\langle B\right\rangle _{t}=dK_{t}.
\label{v.7.7}
\end{equation}

\begin{theorem}
\label{v.th5.1}For state-dependent payoffs, the maximal instantaneous P\&L
of being short a delta-hedged option is of the form \eqref{v.7.7}.
\end{theorem}

\textsc{Proof}: We consider the risk-neutral \& mean-certain world. The
stock price follows%
\begin{equation}
dS_{t}=S_{t}\left( r_{t}dt+dB_{t}\right) ,  \label{v.7.8}
\end{equation}%
where $\left( r_{t}\right) $ is assumed to be a bounded function. Let $V$ be
the unique smooth solution of Barenblatt equation \eqref{v.3.10}. Then by It%
\^{o}'s formula,

\begin{equation}
dV(S_{t})=\frac{\partial V}{\partial t}dt+S_{t}\frac{\partial V}{\partial S}%
\left( r_{t}dt+dB_{t}\right) +\frac{1}{2}S_{t}^{2}\frac{\partial ^{2}V}{%
\partial S^{2}}d\left\langle B\right\rangle _{t}.  \label{v.7.9}
\end{equation}%
The discrete versions of equations \eqref{v.7.8} and \eqref{v.7.9} are
\begin{equation}
\Delta S_{t}=S_{t}\left( r_{t}\Delta t+\Delta B_{t}\right)  \label{v.7.10}
\end{equation}%
and%
\begin{equation}
\Delta V=\frac{\partial V}{\partial t}\Delta t+S_{t}\frac{\partial V}{%
\partial S}\left( r_{t}\Delta t+\Delta B_{t}\right) +\frac{1}{2}S_{t}^{2}%
\frac{\partial ^{2}V}{\partial S^{2}}\Delta \left\langle B\right\rangle _{t}.
\label{v.7.11}
\end{equation}

For a delta-hedging portfolio $\Pi $, the holder of this portfolio is short
one derivative and long an amount $\frac{\partial V}{\partial S}$ of shares
of stocks and $\left( V-\frac{\partial V}{\partial S}S\right) $ cash left in
a bank account. Namely the P\&L variance of the portfolio is
\begin{equation}
\Delta \Pi _{t}=\frac{\partial V}{\partial S}\Delta S_{t}-\Delta V+\left( V-%
\frac{\partial V}{\partial S}S_{t}\right) r_{t}\Delta t.  \label{v.7.12}
\end{equation}%
Now, substituting equations \eqref{v.7.10} and \eqref{v.7.11} into %
\eqref{v.7.12} yields%
\begin{equation*}
\Delta \Pi _{t}=-\frac{\partial V}{\partial t}\Delta t-\frac{1}{2}S_{t}^{2}%
\frac{\partial ^{2}V}{\partial S^{2}}\Delta \left\langle B\right\rangle
_{t}+\left( V-\frac{\partial V}{\partial S}S_{t}\right) r_{t}\Delta t.
\end{equation*}%
Moreover, as the superhedging price of the option follows the Barenblatt
equation \eqref{v.3.10}, we get \
\begin{equation*}
\Delta \Pi _{t}=\frac{1}{2}\underset{\underline{\sigma }\leq \sigma _{t}\leq
\overline{\sigma }}{\sup }\{\sigma _{t}^{2}S_{t}^{2}\Gamma _{t}\}\Delta t-%
\frac{1}{2}S_{t}^{2}\Gamma _{t}\Delta \left\langle B\right\rangle _{t}.
\end{equation*}%
Hence the final P\&L on $(t,t+dt)$ reads
\begin{equation*}
P\&L_{(t,t+dt)}=\frac{1}{2}\underset{\underline{\sigma }\leq \sigma _{t}\leq
\overline{\sigma }}{\sup }\{\sigma _{t}^{2}S_{t}^{2}\Gamma _{t}\}dt-\frac{1}{%
2}S_{t}^{2}\Gamma _{t}d\left\langle B\right\rangle _{t}.
\end{equation*}
$\Box $

Therefore $K_{t}$ over $(t,t+dt)$ coincides with the maximal P\&L of being
short a delta-hedged option. That is, by choosing appreciate managing
volatility $\sigma $, we obtain a nonnegative P\&L (or $K$) for a robust
strategy. Then we come back to equality \eqref{v.4.4} in section \ref%
{v.sect4}, which now has a clear meaning that:

$\bullet $ The minimal superstrategy satisfies: changes of values of the
portfolio minus the instantaneous P\&L, equals to the change of the managing
price of the option. That is to say, we can withdraw money $P\&L_{(t,t+dt)}$
along the way and end up with the terminal payoff.

For option buyers, to guarantee a profit, he/she has to choose the minimal
volatility such that his/her P\&L on $(t,t+dt)$
\begin{equation}
P\&L_{(t,t+dt)}=\frac{1}{2}S_{t}^{2}\widetilde{\Gamma }_{t}d\left\langle
B\right\rangle _{t}-\frac{1}{2}\underset{\underline{\sigma }\leq \sigma
_{t}\leq \overline{\sigma }}{\inf }\{\sigma _{t}^{2}S_{t}^{2}\widetilde{%
\Gamma }_{t}\}dt
\end{equation}%
will always be nonnegative.

\subsection{Estimating the spread}

\label{v.sect5.4}Considering the following minimal superstrategy%
\begin{align*}
dV_{t}& =r_{t}V_{t}dt+\pi _{t}dB_{t}-\left( 2G\left( \eta _{t}\right)
dt-\eta _{t}d\left\langle B\right\rangle _{t}\right) , \\
V_{T}& =\Phi (S_{T})
\end{align*}%
and maximal substrategy%
\begin{align*}
d\widetilde{V}_{t}& =r_{t}\widetilde{V}_{t}dt+\widetilde{\pi }%
_{t}dB_{t}+\left( 2G\left( \widetilde{\eta }_{t}\right) dt-\widetilde{\eta }%
_{t}d\left\langle B\right\rangle _{t}\right) , \\
\widetilde{V}_{T}& =\Phi (S_{T}),
\end{align*}%
where $S$ is defined by \eqref{v.7.8}, $\eta _{t}=\frac{1}{2}S_{t}^{2}\Gamma
_{t}$, and $\widetilde{\eta }_{t}=-\frac{1}{2}S_{t}^{2}\widetilde{\Gamma }%
_{t}$. In an incomplete market the superhedging price and subhedging price
(also called ask/bid price) do not usually equal to each other and a set of
hedging prices exist. \textrm{\cite{cont} }proposed to measure the impact of
model uncertainty on the value of a contingent claim $\xi $ by
\begin{equation*}
e_{\mathcal{P}}(\xi ):=V_{0}\left[ \xi \right] -\widetilde{V}_{0}\left[ \xi %
\right] .
\end{equation*}%
Define $D_{t}={\normalsize \exp }\left\{ -{\normalsize \int_{0}^{t}r_{s}ds}%
\right\} $. Since $e_{\mathcal{P}}(\xi )=\mathbb{E}\left[ D_{T}\xi \right] +%
\mathbb{E}\left[ -D_{T}\xi \right] $, then $e_{\mathcal{P}}(\cdot )$
satisfies

(i) $e_{\mathcal{P}}(D_{T}^{-1}c)=0,\forall c\in \mathbf{R}$,

(ii) $e_{\mathcal{P}}(\xi +\eta )\leq e_{\mathcal{P}}(\xi )+e_{\mathcal{P}%
}(\eta ),\forall \xi ,\eta \in L{_{G}^{\mathrm{p}}}(\Omega )$, $p>1$,

(iii) $e_{\mathcal{P}}(\xi )\geq 0$.

The following result shows that $e_{\mathcal{P}}(\cdot)$ depends on closely
the volatility uncertainty and gamma risk.

\begin{theorem}
\label{v.th5.2}For all $\xi =\Phi (S_{T})$, $\Phi $ is a Lipschitz function
of $S_{T}$, we have%
\begin{equation}
e_{\mathcal{P}}(\xi )\leq \left( \overline{\sigma }^{2}-\underline{\sigma }%
^{2}\right) \cdot L,
\end{equation}%
where $L=\mathbb{E}\left[ \int_{0}^{T}D_{t}S_{t}^{2}\max (\left\vert \Gamma
_{t}\right\vert ,|\widetilde{\Gamma }_{t}|)dt\right] $.
\end{theorem}

\textsc{Proof}: \ We denote $\overline{V}_{t}=V_{t}-\widetilde{V}_{t}$, $%
\overline{\pi }_{t}=\pi _{t}-\widetilde{\pi }_{t}$, $\overline{K}%
_{t}=\int_{0}^{t}2G\left( \eta _{s}\right) ds-\int_{0}^{t}\eta
_{s}d\left\langle B\right\rangle _{s}+\int_{0}^{t}2G\left( \widetilde{\eta }%
_{s}\right) ds-\int_{0}^{t}\widetilde{\eta }_{s}d\left\langle B\right\rangle
_{s}$. Then%
\begin{equation*}
\overline{V}_{t}=0-\int_{t}^{T}r_{s}\overline{V}_{s}ds-\int_{t}^{T}\overline{%
\pi }_{s}dB_{s}-\int_{t}^{T}d\overline{K}_{s}.
\end{equation*}%
Note that in general $\left( \overline{K}_{s}\right) $ is not a G-martingale
since $G$ is a subadditive function. Applying It\^{o}'s formula to ($D_{t}%
\overline{V}_{t}$) we get%
\begin{equation}
\overline{V}_{t}=D_{t}^{-1}\mathbb{E}\left[ \int_{t}^{T}D_{s}d\overline{K}%
_{s}|\mathcal{F}_{t}\right] ,q.s.  \label{v.7.15}
\end{equation}%
Therefore%
\begin{align*}
e_{\mathcal{P}}(\xi )& =\overline{V}_{0}=\mathbb{E}\left[ \int_{0}^{T}D_{s}d%
\overline{K}_{s}\right] \\
& \leq \left( \overline{\sigma }^{2}-\underline{\sigma }^{2}\right) \cdot
\mathbb{E}\left[ \int_{0}^{T}D_{t}(\left\vert \eta _{t}\right\vert
+\left\vert \widetilde{\eta }_{t}\right\vert )dt\right] \\
& \leq \frac{1}{2}\left( \overline{\sigma }^{2}-\underline{\sigma }%
^{2}\right) \cdot \mathbb{E}\left[ \int_{0}^{T}D_{t}S_{t}^{2}(\left\vert
\Gamma _{t}\right\vert +|\widetilde{\Gamma }_{t}|)dt\right] \\
& \leq \left( \overline{\sigma }^{2}-\underline{\sigma }^{2}\right) \cdot
\mathbb{E}\left[ \int_{0}^{T}D_{t}S_{t}^{2}\max (\left\vert \Gamma
_{t}\right\vert ,|\widetilde{\Gamma }_{t}|)dt\right] .
\end{align*}%
$\Box $

Theorem \ref{v.th5.2} and results in Section \ref{v.sect5.3} also hold for
discrete-path-dependent payoffs.

\begin{remark}
Observing from \eqref{v.7.15} that, the ask-bid spread is in fact the
accumulation of summation of the superhedging $P\&L$ and the subhedging $%
P\&L $.
\end{remark}

\section{Conclusion}

\label{v.sect6}We consider mean-volatility uncertainty by Peng's
G-stochastic analysis in this paper. All results can be applied to
path-dependent options. Price of stock is assumed to be generalized
geometric G-Brownian motion in which the mean-uncertainty is not necessarily
related to the volatility-uncertainty. A neat formulation of superhedging
problem is given by BSDE driven by G-Brownian motion. For subhedging we have
to impose strong conditions to guarantee no-arbitrage, which is essentially
different from Vorbink's work.

Another phenomenon deserving mention is that the mean-uncertainty does not
influence pricing a security. When we deriving the superhedging PDEs, the
stock appreciation disappears after delta-hedging, which shows that there is
a risk-neutral world under which all investors price and hedge in a
risk-neutral \& mean-certain way.

In Markovian setting, we give a precise and practical explanation of the
finite-variance term in the minimal superstrategy in the language of P\&L.
The control of price fluctuations by volatility interval are also discussed.

All shows that G-stochastic analysis is a convenient tool to measure model
uncertainty. Although, in the eloquent words of \textrm{\cite{der}}: even
the finest model is only a model of the phenomena, and not the real thing,
we believe we are modeling in a more efficient way to solve problems of the
real thing.

\appendix

\section{Peng's G-stochastic calculus}

\label{v.appA}In this section we recall some necessary notions and lemmas of
Peng's G-stochastic calculus needed in this paper. Readers could refer to
\textrm{\cite{p10a}} for more systematic information.

For two stochastic processes $(X_{t})$ and $(Y_{t})$, let $\left\langle
X,Y\right\rangle _{t}$ denote their mutual variance. We denote by $\mathbb{S}%
(n)$ the collection of $n\times n$ symmetric matrices, $\mathbb{S}_{+}(d)$
the positive-semidefinite elements of $\mathbb{S}(d)$. We observe that $%
\mathbb{S}(n)$ is a Euclidean space with the scalar product $\left\langle
A,B\right\rangle =tr[AB]$. Let $\Omega $ be a complete metrizable and
separable space. Typically we can take $\Omega =C_{0}([0,+\infty ),\mathbf{R}%
^{d})$ with the topology of uniform convergence on compact subspaces. $%
\mathcal{B}(\Omega )$ denotes the Borel $\sigma $-algebra of $\Omega $. Let $%
{\mathcal{H}}$ be a linear space of real functions defined on $\Omega $ such
that if $X_{1},\ldots ,X_{n}\in {\mathcal{H}}$ then $\varphi (X_{1},\ldots
X_{n})\in {\mathcal{H}}$ for each $\varphi \in C_{l.Lip(\mathbf{R}^{n})}$
where $C_{l.Lip(\mathbf{R}^{n})}$ denotes the linear space of (local
Lipschitz) functions $\varphi $ satisfying
\begin{equation*}
|\varphi \left( x\right) -\varphi \left( y\right) |\leq C(1+\left\vert
x\right\vert ^{m}+\left\vert y\right\vert ^{m})|x-y|,\ \forall x,y\in
\mathbf{R}^{n},
\end{equation*}%
for some $C>0$, $m\in N$ depending on $\varphi $. ${\mathcal{H}}$ is
considered as a space of `random variables'. In this case $X=(X_{1},\ldots
,X_{n})$ is called an $n$-dimensional random vector, denoted by $X\in
\mathcal{H}^{n}$. We also denote by $C_{b}^{k}(\mathbf{R}^{n})$ the space of
bounded and $k$-time continuously differentiable functions with bounded
derivatives of all orders less than or equal to $k$; $C_{Lip(\mathbf{R}%
^{n})} $ the space of Lipschitz continuous functions.

\begin{definition}
A \textbf{sublinear expectation} $\mathbb{E}$ on ${\mathcal{H}}$ is a
functional $\mathbb{E}:{\mathcal{H\mapsto}}\mathbf{R}$ satisfying the
following properties: for all $X,Y\in{\mathcal{H}}$, we have

(a) Monotonicity: If $X\geq Y$, then $\mathbb{E[}X\mathbb{]}\geq \mathbb{E[}Y%
\mathbb{]}$.

(b) Constant preserving: $\mathbb{E[}c\mathbb{]=}c,\forall c\in\mathbf{R}$.

(c) Sub-additivity: $\mathbb{E[}X+Y\mathbb{]}\leq\mathbb{E[}X\mathbb{]}+%
\mathbb{E[}Y\mathbb{]}$.

(d) Positive homogeneity: $\mathbb{E[\lambda}X\mathbb{]=\lambda E[}X\mathbb{]%
},\forall\lambda\geq0$.
\end{definition}

\begin{definition}
Let $X_{1}$ and $X_{2}$ be two n-dimensional random vectors defined on
nonlinear expectation spaces $(\Omega _{1},\mathcal{H}_{1},\mathbb{E}_{1})$
and $(\Omega _{2},\mathcal{H}_{2},\mathbb{E}_{2})$ respectively. They are
called identically distributed, denoted by $X_{1}\overset{d}{=}X_{2}$, if%
\begin{equation*}
\mathbb{E}_{1}[\varphi (X_{1})]=\mathbb{E}_{2}[\varphi (X_{2})],\ \forall
\varphi \in C_{l.Lip}(\mathbf{R}^{n}).
\end{equation*}
\end{definition}

\begin{definition}
In a sublinear expectation space $(\Omega ,\mathcal{H},\mathbb{E})$ a random
vector $Y\in {\mathcal{H}}^{n}$ is said to be independent of another random
vector $X\in {\mathcal{H}}^{m}$ under $\mathbb{E}$ if for each test function
$\varphi \in C_{l.Lip}(\mathbf{R}^{m+n})$ we have%
\begin{equation*}
\mathbb{E}[\varphi (X,Y)]=\mathbb{E}\left[ \mathbb{E}[\varphi (x,Y)]_{x=X}%
\right] .
\end{equation*}
\end{definition}

\begin{remark}
It is interesting that $Y$ is independent of $X$does not necessarily imply $%
X $ is independent of $Y$. See Chapter I, Example 3.13 in \textrm{\cite{p10a}%
.}
\end{remark}

\begin{definition}
($G$\textbf{-normal distribution)}. A d-dimensional random vector $%
X=(X_{1},...,X_{d})$ in a sublinear expectation space $(\Omega ,\mathcal{H},%
\mathbb{E})$ is called G-normal distributed if for each $a,b>0$ we have%
\begin{equation*}
aX+b\overline{X}\,\overset{d}{=}\sqrt{a^{2}+b^{2}}X
\end{equation*}%
where $\overline{X}$ is an independent copy of $X$.
\end{definition}

\begin{remark}
It is easy to check that $\mathbb{E}[X]=$ $\mathbb{E}[-X]=0$. The so called
`G' is related to $G:\mathbb{S}(d)\mapsto \mathbf{R}$ defined by
\begin{equation*}
G\left( A\right) =\frac{1}{2}\mathbb{E[}\left\langle AX,X\right\rangle
\mathbb{]},
\end{equation*}
\end{remark}

\textrm{\cite{hp}} proved that for a sublinear expectation $\mathbb{E}$ on $%
(\Omega,\mathcal{H})$, there exists a family of linear expectation \{$%
E_{P};P\in\mathcal{P}$\} on $(\Omega,\mathcal{H})$ such that $\mathbb{E}%
\left[ \cdot\right] =\underset{P\in\mathcal{P}}{\sup }{\normalsize E}_{P}%
\left[ \cdot\right] $.

\begin{definition}
For a given set of probability measures $\mathcal{P}$, we introduce the
natural Choquet capacity
\begin{equation*}
C(A):=\underset{P\in \mathcal{P}}{\sup }P(A),\ A\in \mathcal{B}(\Omega ).
\end{equation*}%
A property holds quasi-surely(q.s.) if it holds outside a polar set A, i.e.,
$C(A)=0$. A mapping $X$ on $\Omega $ with values in a topological space is
said to be quasi-continuous (q.c.) if $\forall \varepsilon >0$, there exists
an open set $O$ with $C(O)<\varepsilon $ such that $X|_{O^{c}}$ is
continuous.
\end{definition}

\begin{definition}
(\textbf{G--Brownian motion}). A d-dimensional process $(B_{t})_{t\geq0}$ on
a sublinear expectation space $(\Omega,\mathcal{H},\mathbb{E})$ is called a
G--Brownian motion if the following properties are satisfied:

(i) $B_{0}(\omega)=0$;

(ii) For each $t,s\geq0$, the increment $B_{t+s}-B_{t}$ is independent from $%
(B_{t_{1}},B_{t_{2}},\ldots,B_{t_{n}})$, for each $n\in N$ and $0\leq
t_{1}\leq\cdots\leq t_{n}\leq t$;

(iii) $B_{t+s}-B_{t}\overset{d}{=}\sqrt{s}X$, where $X$ is $G$-normal
distributed.
\end{definition}

\begin{definition}
(\textbf{Maximal distribution)}. A d-dimensional random vector $%
X=(X_{1},...,X_{d})$ in a sublinear expectation space $(\Omega,\mathcal{H},%
\mathbb{E})$ is called maximal distributed if for each $a,b>0$ we have%
\begin{equation*}
aX+b\overline{X}\,\overset{d}{=}\left( a+b\right) X
\end{equation*}
where $\overline{X}$ is an independent copy of $X$.
\end{definition}

\begin{remark}
For a maximal distributed random variable $X$, there exists a bounded,
closed and convex subset $\Gamma\in\mathbf{R}^{d}$ such that
\begin{equation*}
\mathbb{E}[\varphi(X)]=\underset{a\in\Gamma}{\max}\varphi(a),\ \forall
\varphi\in C_{l.Lip}(\mathbf{R}^{d}).
\end{equation*}
\end{remark}

\begin{definition}
(\textbf{Finite-variance G--Brownian motion}). A d-dimensional process $%
(\beta _{t})_{t\geq 0}$ on a sublinear expectation space $(\Omega ,\mathcal{H%
},\mathbb{E})$ is called a finite-variance G--Brownian motion if the
following properties are satisfied:

(i) $\beta _{0}(\omega )=0$;

(ii) For each $t,s\geq 0$, the increment $\beta _{t+s}-\beta _{t}$ is
independent from $(\beta _{t_{1}},\beta _{t_{2}},\ldots ,\beta _{t_{n}})$,
for each $n\in N$ and $0\leq t_{1}\leq \cdots \leq t_{n}\leq t$;

(iii) $\beta _{t+s}-\beta _{t}\overset{d}{=}sX$, where $X$ is maximal
distributed.
\end{definition}

Typically, $\left\langle B\right\rangle _{t}$, the quadratic variance
process of $\left( B_{t}\right) $, is a finite-variance G--Brownian motion.
We conclude properties of finite-variance G--Brownian motion as following

\begin{proposition}
Let $(\beta _{t})$ be a one-dimensional finite-variance G--Brownian motion.
Then

(i) $(\beta _{t})$ is a continuous process with finite variance, independent
and stationary increments under $\mathbb{E}$.

(ii)
\begin{equation}
\mathbb{E}[\varphi (\beta _{t+s}-\beta _{s})|\mathcal{F}_{s}]=\underset{%
\underline{\mu }\leq \mu \leq \overline{\mu }}{\max }\varphi (\mu t),\
\forall \varphi \in C_{l.Lip}(\mathbf{R}).
\end{equation}%
where we denote the usual parameters $\overline{\mu }=\mathbb{E}[\beta _{1}]$%
, $\underline{\mu }=-\mathbb{E}[-\beta _{1}]$.

(iii) For each $0\leq t\leq T<\infty $, we have $q.s.$
\begin{equation}
\underline{\mu }(T-t)\leq \beta _{T}-\beta _{t}\leq \overline{\mu }(T-t).
\end{equation}
\end{proposition}

\textsc{Proof}: \ See \textrm{\cite{p10a}} for (i), (ii) and (iii). $\Box $

In the sequence, let $\Omega =C_{0}([0,+\infty ),\mathbf{R}^{d})$ denote the
space of all $\mathbf{R}^{d}-$valued continuous paths $(\omega _{t})_{t\in
R^{+}}$ with $\omega _{0}=0$, by $C_{b}(\Omega )$ all bounded and continuous
functions on $\Omega $. For each fixed $T\geq 0$, we consider the following
space of random variables:%
\begin{equation*}
L_{ip}(\Omega _{T}):=\{X(\omega )=\varphi (\omega _{t_{1}\wedge
T},...,\omega _{t_{m}\wedge T}),\forall m\geq 1,\forall \varphi \in C_{l.Lip(%
\mathbb{R}^{m})}\}.
\end{equation*}%
We also denote
\begin{equation*}
L_{ip}(\Omega ):=\overset{\infty }{\underset{n=1}{\cup }}L_{ip}(\Omega _{n}).
\end{equation*}%
We will consider the canonical space and set $B_{t}(\omega )=\omega _{t}$.
For a given sublinear function $G\left( A\right) =\frac{1}{2}\underset{%
\gamma \in \Gamma }{\sup }\left\{ tr[A\gamma ]\right\} $, where $A\in
\mathbb{S}(d)$, $\Gamma $ is a given nonempty, bounded and closed convex
subset of $\mathbb{S}_{+}(d)$, by the following%
\begin{equation*}
\partial _{t}u(t,x)-G\left( D_{x}^{2}u\right) =0,\ u(0,x)=\varphi (x),
\end{equation*}%
\textrm{\cite{p06} }defined $G$-expectation $\mathbb{E}$ as $\mathbb{E}%
[\varphi (x+B_{t})]=u(t,x)$. For each $p\geq 1$, $X\in L_{ip}(\Omega )$, $%
\Vert X\Vert _{p}=\left( \mathbb{E}[\left\vert X\right\vert ^{p}]\right) ^{%
\frac{1}{p}}$ forms a norm and $\mathbb{E}$ can be continuously extended to
a Banach space, denoted by $L{_{G}^{\mathrm{p}}}(\Omega )$. \textrm{\cite{hp}
}proved that $L{_{G}^{\mathrm{p}}}(\Omega )=\{X|\ X$ is $\mathcal{B}(\Omega
)-$measurable and has a quasi-continuous version, s.t. $\underset{%
n\rightarrow \infty }{\lim }\mathbb{E}[\left\vert X\right\vert
^{p}1_{\left\{ \left\vert X\right\vert >n\right\} }]=0\}$. By the method of
Markov chains, \textrm{\cite{p06,p08}} also defined corresponding
conditional expectation, $\mathbb{E}\left[ \cdot |\Omega _{t}\right] :L{%
_{G}^{\mathrm{1}}}(\Omega )\mapsto L{_{G}^{\mathrm{1}}}(\Omega _{t})$,%
 where $\Omega _{t}:=\left\{
\omega ._{\wedge t}:\omega \in \Omega \right\} $. Under $\mathbb{E}\left[
\cdot \right] $, the canonical process $B_{t}(\omega )=\omega _{t}$, $t\in
\lbrack 0,\infty )$ is a $G$-Brownian motion.

The following properties hold for $\mathbb{E}\left[ \cdot|\Omega_{t}\right] $
\textit{q.s.}.

\begin{proposition}
For $X,Y\in L{_{G}^{\mathrm{1}}}(\Omega)$, we have q.s.,

(i) $\mathbb{E}[\eta X|\Omega_{t}]=\eta^{+}\mathbb{E}[X|\Omega_{t}]+\eta ^{-}%
\mathbb{E}[-X|\Omega_{t}]$, for bounded $\eta\in L{_{G}^{\mathrm{1}}}%
(\Omega_{t})$.

(ii) If $\mathbb{E}[X|\Omega_{t}]=-\mathbb{E}[-X|\Omega_{t}]$, for some $t$,
then $\mathbb{E[}X+Y|\Omega_{t}\mathbb{]}=\mathbb{E[}X|\Omega_{t}\mathbb{]}+%
\mathbb{E[}Y|\Omega_{t}\mathbb{]}$.

(iii) $\mathbb{E[}X+\eta|\Omega_{t}\mathbb{]}=\mathbb{E[}X|\Omega _{t}%
\mathbb{]}+\eta$, $\eta\in L{_{G}^{\mathrm{1}}}(\Omega_{t})$.
\end{proposition}

For a partition of $[0,T]$: $0=t_{0}<t_{1}<\cdots <t_{N}=T$ and $p\geq 1$,
we set

${\mathcal{M}_{G}^{\mathrm{p,0}}}(0,{\mathit{T}})$: the collection of
processes $\eta _{t}(\omega )=\sum_{j=0}^{N}\xi _{j}(\omega )\cdot
1_{[t_{j},t_{j+1}]}(t)$, where $\xi _{j}\in L{_{G}^{\mathrm{p}}}(\Omega
_{t_{j}}),j=0,1,...,N$;

$\mathcal{M}{_{G}^{\mathrm{P}}}(0,{\mathit{T}})$: the completion of ${%
\mathcal{M}_{G}^{\mathrm{p,0}}}(0,{\mathit{T}})$ under norm $||\eta ||_{%
\mathcal{M}}=\left( \mathbb{E}\left[ \int_{0}^{T}|\eta_{t}|^{p}dt\right]
\right) ^{^{\frac{1}{p}}}$;

$\mathcal{H}{_{G}^{\mathrm{P}}}(0,{\mathit{T}})$: the completion of ${%
\mathcal{M}_{G}^{\mathrm{p,0}}}(0,{\mathit{T}})$ under norm $||\eta ||_{%
\mathcal{H}}=\left( \mathbb{E}\left( \int_{0}^{T}|\eta_{t}|^{2}dt\right) ^{%
\frac{p}{2}}\right) ^{\frac{1}{p}}$. It is easy to prove that $\mathcal{H}{%
_{G}^{\mathrm{2}}}(0,{\mathit{T}})=\mathcal{M}{_{G}^{\mathrm{2}}}(0,{\mathit{%
T}})$.

For any $\left( \eta _{t}\right) \in \mathcal{M}{_{G}^{\mathrm{2}}}$, G-It%
\^{o} integral is well defined in \textrm{\cite{p06,p08} }and extended to $%
\mathcal{H}{_{G}^{\mathrm{P}}}$ by \textrm{\cite{song}}.

\section{BSDE with linear generator and driven by G-Brownian motion}

\label{v.appB}We define $D_{t}={\normalsize \exp }\left\{ -{\normalsize %
\int_{0}^{t}r_{s}ds}\right\} $. Then $D_{t}$ satisfies%
\begin{equation*}
dD_{t}=-D_{t}r_{t}dt\text{, and }D_{t=0}=1.
\end{equation*}

Consider the following one dimensional BSDE with linear generator and driven
by one dimensional G-Brownian motion:%
\begin{align}
dY_{t}& =\left( r_{t}Y_{t}-\phi _{t}\right) dt+Z_{t}dB_{t}-dK_{t},\
\label{v.C.1} \\
Y_{T}& =\xi ,  \notag
\end{align}%
where $\xi \in L{_{G}^{\mathrm{p}}}(\Omega _{T})$, $p>1$, $r_{t}$ and $\phi
_{t}$ are $\mathcal{F}_{t}$-measurable bounded processes belonging to ${%
\mathcal{M}_{G}^{\mathrm{p}}}$.

\begin{definition}
A solution to BSDE \eqref{v.C.1} is a triple of adapted processes $%
(Y_{t},Z_{t},K_{t})$ where $(K_{t})$ is a continuous, increasing process
with $K_{0}=0$ and $(-K_{t})$ being a G-martingale.
\end{definition}

For BSDE \eqref{v.C.1}, we have

\begin{theorem}
\label{v.thB.1}There is a unique triple $(Y_{t},Z_{t},K_{t})$ satisfying %
\eqref{v.C.1} with $Y\in {\mathcal{M}_{G}^{\mathrm{p}}}${, }$Z\in \mathcal{H}%
{_{G}^{\mathrm{\alpha }}}$ and $K_{T}\in L{_{G}^{\mathrm{\alpha }}}(\Omega
_{T})$, $1\leq \alpha <p$, $p>1$. Furthermore we have q.s.,%
\begin{equation}
Y_{t}=D_{t}^{-1}\mathbb{E}\left[ D_{T}\xi +\int_{t}^{T}D_{s}\phi _{s}ds|%
\mathcal{F}_{t}\right] .\   \label{v.C.2}
\end{equation}
\end{theorem}

\textsc{Proof}: \ Consider the following BSDE under sublinear expectation $%
\mathbb{E}$:
\begin{equation}
Y_{t}=\mathbb{E}\left[ \xi -\int_{t}^{T}\left( r_{s}Y_{s}-\phi _{s}\right)
ds|\mathcal{F}_{t}\right] .  \label{v.C.3}
\end{equation}%
By the technique of contracting mapping principle employed in \textrm{\cite%
{p10a}, Ch.V, Sec. 2, }one can similarly prove that there is a unique
solution $Y\in {\mathcal{M}_{G}^{\mathrm{p}}}$ to BSDE \eqref{v.C.3}.
Applying martingale representation theorem established in \textrm{\cite{song}%
}, there is a unique pair $(Z,K)$ with $Z\in {\mathcal{M}_{G}^{\mathrm{%
\alpha }}}$ and $K_{T}\in L{_{G}^{\mathrm{\alpha }}}(\Omega _{T})$, $1\leq
\alpha <p$ such that%
\begin{equation*}
\mathbb{E}\left[ \xi -\int_{0}^{T}\left( r_{s}Y_{s}-\phi _{s}\right) ds|%
\mathcal{F}_{t}\right] =Y_{0}+\int_{0}^{t}Z_{s}dB_{s}-K_{t},\ \mathcal{P-}%
q.s.
\end{equation*}%
Hence
\begin{align*}
Y_{t}& =\mathbb{E}\left[ \xi -\int_{0}^{T}\left( r_{s}Y_{s}-\phi _{s}\right)
ds|\mathcal{F}_{t}\right] +\int_{0}^{t}\left( r_{s}Y_{s}-\phi _{s}\right) ds
\\
& =Y_{0}++\int_{0}^{t}\left( r_{s}Y_{s}-\phi _{s}\right)
ds+\int_{0}^{t}Z_{s}dB_{s}-K_{t},\ \mathcal{P-}q.s.
\end{align*}%
or in backward form%
\begin{equation*}
Y_{t}=\xi -\int_{t}^{T}\left( r_{s}Y_{s}-\phi _{s}\right)
ds-\int_{t}^{T}Z_{s}dB_{s}+\int_{t}^{T}dK_{s},\ \mathcal{P-}q.s.
\end{equation*}%
Thus the triple $(Y_{t},Z_{t},K_{t})$ constructed by above procedure is a
solution of \eqref{v.C.1}.

Applying It\^{o}'s formula to $D_{t}Y_{t}$, we have
\begin{align*}
d(D_{t}Y_{t})& =D_{t}\left[ \left( r_{t}Y_{t}-\phi _{t}\right)
dt+Z_{t}dB_{t}-dK_{t}\right] -Y_{t}D_{t}r_{t}dt \\
& =-D_{t}\phi _{t}dt+D_{t}Z_{t}dB_{t}-D_{t}dK_{t}.
\end{align*}%
Note that $(D_{t}Y_{t}+\int_{0}^{t}D_{s}\phi _{s}ds)$ is a G-martingale.
Hence%
\begin{align*}
D_{t}Y_{t}& =\mathbb{E}\left[ D_{T}\xi +\int_{0}^{T}D_{s}\phi _{s}ds|%
\mathcal{F}_{t}\right] -\int_{0}^{t}D_{s}\phi _{s}ds \\
& =\mathbb{E}\left[ D_{T}\xi +\int_{t}^{T}D_{s}\phi _{s}ds|\mathcal{F}_{t}%
\right] .
\end{align*}%
Therefore the solution of \eqref{v.C.1} has the following unique form:
\begin{equation*}
Y_{t}=D_{t}^{-1}\mathbb{E}\left[ D_{T}\xi +\int_{t}^{T}D_{s}\phi _{s}ds|%
\mathcal{F}_{t}\right] .
\end{equation*}

$\Box$

Let $Y^{i}$ be the solution of \eqref{v.C.1} with parameters $\left( \xi
^{i},\phi ^{i}\right) $, $i=1,2$. It is interesting that ($Y^{1}+Y^{2}$) is
no longer a solution of \eqref{v.C.1} with parameters $\left( \xi ^{1}+\xi
^{2},\phi ^{1}+\phi ^{2}\right) $, though BSDE \eqref{v.C.1} has a linear
generator. All attributes to the sublinearity. $(-K_{t}^{1}-K_{t}^{2})$ is
no more a G-martingale. We have the following

\begin{corollary}
Let $\widetilde{Y}$ be the solution of \eqref{v.C.1} with parameters $\left(
\xi^{1}+\xi^{2},\phi^{1}+\phi^{2}\right) $. Then
\begin{equation*}
Y^{1}+Y^{2}\geq\widetilde{Y}.
\end{equation*}
\end{corollary}

\textsc{Proof}: \ It is just a sequence of \eqref{v.C.2} and the
sublinearity of G-expectation $\mathbb{E}$. $\Box $

This property reflects that if two agent cooperate with each other, then
superhedging the whole might yield less pricing error.

\end{document}